\documentclass[fleqn,usenatbib]{mnras} 

\usepackage{newtxtext,newtxmath}

\usepackage[T1]{fontenc}
\usepackage{ae,aecompl}

\usepackage{graphicx}	
\usepackage{amsmath}	
\usepackage{amssymb}	
\usepackage{grffile}    

\def\km{{\rm\thinspace km}}
\def\s{{\rm\thinspace s}}

\def\kmps{\hbox{${\rm\km\s^{-1}\,}$}}

\def\Msol{\hbox{${\rm\thinspace M_{\odot}}$}}

\def\spose#1{\hbox to 0pt{#1\hss}}
\def\ltsimm{\mathrel{\spose{\lower 3pt\hbox{$\sim$}}
        \raise 2.0pt\hbox{$<$}}}
\def\gtsimm{\mathrel{\spose{\lower 3pt\hbox{$\sim$}}
        \raise 2.0pt\hbox{$>$}}}

\title[Mass-loaded supernova remnant momentum and energy injection]
{Momentum and energy injection by a supernova remnant into an inhomogeneous medium}

\author[J.~M.~Pittard]
{J.~M.~Pittard\thanks{E-mail: j.m.pittard@leeds.ac.uk (JMP)}\\
School of Physics and Astronomy, University of
       Leeds, Woodhouse Lane, Leeds LS2 9JT, UK
}

\date{Accepted 2019 July 05. Received 2019 June 12; in original form 2019 May 09}

\pubyear{2019}

\begin{document}
\label{firstpage}
\pagerange{\pageref{firstpage}--\pageref{lastpage}}
\maketitle

\begin{abstract}
We investigate the effect of mass-loading from embedded clouds on the evolution of supernova remnants and on the energy and momentum that they inject into an inhomogeneous interstellar medium. We use 1D hydrodynamical calculations and assume that the clouds are numerous enough that they can be treated in the continuous limit. The destruction of embedded clouds adds mass into the remnant, increasing its density and pressure, and decreasing its temperature. The remnant cools more quickly, is less able to do PdV work on the swept-up gas, and ultimately attains a lower final momentum (by up to a factor of two or more). We thus find that the injection of momentum is more sensitive to an inhomogeneous environment than previous work has suggested, and we provide fits to our results for the situation where the cloud mass is not limited. The behaviour of the remnant is more complex in situations where the cloud mass is finite and locally runs out. In the case of multiple supernovae in a clustered environment, later supernova explosions may encounter higher densities than previous explosions due to the prior liberation of mass from engulfed clouds. If the cloud mass is finite, later explosions may be able to create a sustained hot phase when earlier explosions have not been able to.
\end{abstract}

\begin{keywords}
galaxies: evolution -- galaxies: formation -- galaxies: ISM -- supernovae: general -- ISM: supernova remnants -- ISM: kinematics and dynamics
\end{keywords}



\section{Introduction}
\label{sec:intro}
Supernova remnants (SNRs) play a key role in determining the nature of the interstellar medium (ISM), and the behaviour and evolution of galaxies. They are responsible for the hot phase of the ISM and for shaping large-scale structures \citep*[e.g.,][]{deAvillez:2004,Hill:2012,Hennebelle:2014,Gatto:2015}; they drive turbulence in the diffuse gas \citep*[e.g.,][]{MacLow:2004,Dib:2006} and in denser molecular clouds \citep{Padoan:2016}; and they generate and maintain turbulent pressure that on large scales limits gravitational condensation and regulates star formation \citep*[e.g.,][]{Ostriker:2011,Shetty:2012}. Overlapping SNRs create galactic winds \citep*[e.g.,][]{Chevalier:1985,Veilleux:2005,Strickland:2009,Hopkins:2012}.

In models of galaxy formation stellar feedback must be included otherwise gas is converted into stars much too efficiently \citep*[e.g.,][]{Hopkins:2011,Teyssier:2013,Agertz:2013}. Without stellar feedback, $5 - 1000\times$ too many stars may be created \citep*[e.g.,][]{White:1991,Keres:2009,Behroozi:2010,Moster:2010,Faucher-Giguere:2011}. Stellar feedback is also necessary to explain the observed distribution of heavy elements in the intergalactic medium \citep*[e.g.,][]{Aguirre:2001,Oppenheimer:2006,Wiersma:2010}. 

Early studies of SNR evolution, and the energy and momentum that they inject into a uniform medium, were made by \citet*{Cioffi:1988} and \cite{Thornton:1998}. The SNR passes through several stages in its evolution: an initial free-expansion ejecta-dominated (ED) stage, an adiabatic Sedov-Taylor (ST) stage, a cooling-modified pressure-driven snowplough (PDS) stage, and a momentum-conserving snowplough (MCS) stage; before finally merging into the ISM \citep[e.g.,][]{Cioffi:1988}. During the ST stage the momentum of the remnant can significantly increase, by an order of magnitude or more, due to the work done by the hot interior gas. However, once the shell forms at the end of the ST stage, the total momentum increases by only another 50 per cent or so. 

Early implementations of supernova feedback in galaxy and cosmological simulations that attempted to use energy-driven feedback suffered from an ``over-cooling'' problem due to insufficient numerical resolution \citep[e.g.,][]{Katz:1992}. Since then many different prescriptions have been trialled, including artificially turning off cooling for a limited period of time \citep*[e.g.,][]{Thacker:2000,Stinson:2006,Governato:2007,Shen:2010,Scannapieco:2012}, or temporarily suppressing hydrodynamical interactions \citep[e.g.,][]{Springel:2003,Oppenheimer:2006,Vogelsberger:2013}. The \citet{Stinson:2006} method of disabling cooling, which has ben widely adopted \citep[e.g.,][]{Guedes:2011,Brook:2012,Stinson:2013}, is now recognised to inject far too much momentum and significantly overestimates the production of hot gas. The \citet{Hopkins:2014} FIRE feedback scheme may also overestimate the final momentum of SNRs \citep{Kimm:2014}.

A further complication is that the ISM is extremely inhomogeneous. Efforts to understand the effect of an inhomogeneous ISM on SNR evolution have been made using numerical simulation \citep{Cowie:1981,Wolff:1987,Arthur:1996,Dyson:2002,Pittard:2003,Korolev:2015,Slavin:2017,Zhang:2019a,Zhang:2019b} and similarity solutions \citep*{McKee:1977,Chieze:1981,Dyson:1987,White:1991b,Chen:1995,Pittard:2001}. A surprising finding from recent 3D hydrodynamical simulations of SNRs expanding into an inhomogeneous medium is that the final momentum injection is remarkably insensitive to the environment, differing by only $5-30$ per cent compared to evolution in a homogeneous environment of the same average density \citep{Iffrig:2015,Kim:2015,Martizzi:2015,Walch:2015}.

In a small minority of galaxy formation simulations there has been some attempt to account for an unresolved inhomogeneous ISM using sub-grid physics \citep[see, e.g.,][and references therein]{Springel:2003,Scannapieco:2006}. More recently, \citet{Kimm:2015} considered an unresolved inhomogeneous ISM in their feedback model NutMFBmp, making the ad-hoc assumption that only 10 per cent of the mass that the SNR encounters is swept-up. This allows their remnants to expand more quickly, as they effectively encounter only gas at the lower inter-cloud density, and drastically reduces the amount of star formation in the simulation. However, no interaction between the overrun clouds and the remnant interior is considered.

In this work we wish to reinvestigate the effect of a clumpy inhomogeneous environment on the evolution and momentum injection of SNRs. All of the 3D calculations that directly model SNR-cloud interactions suffer from insufficient numerical resolution\footnote{\citet[][]{Pittard:2016} show that at least $32-64$ cells per cloud radius is needed to capture the dominant dynamical processes in a shock-cloud interaction, and that poorly resolved clouds accelerate and mix up to $5\times$ faster than they should.}, and the clouds are typically too large and spaced out compared to real ISM clouds. This motivates the current work, where we investigate the mass-loading of remnants where the clouds are treated essentially as continuously distributed mass injection sources. We assume that the clouds are first overrun, and are then destroyed within the remnant interior, and we adopt a constant rate of mass injection per unit volume unless and until the available mass reservoir at a particular radius is exhausted.
Our method obviates the need to model the destruction of individual clouds, which is numerically challenging and physically complicated, while revealing how the destruction of many clouds changes the global properties of the remnant.
This work also improves on earlier work by \citet{Dyson:2002} and \citet{Pittard:2003} by: i) focussing on the final radial momentum; ii) considering the effect of a finite amount of cloud mass to inject into the remnant; and iii) investigating the effect of multiple SNe. In Sec.~\ref{sec:setup} we note the specific details of our calculations. Sec.~\ref{sec:results} presents our results and in Sec.~\ref{sec:discussion} we discuss some implications. We summarize and conclude our work in Sec.~\ref{sec:summary}.

\section{The calculations}
\label{sec:setup}

\subsection{The numerics}
\label{sec:numerics}
The standard inviscid equations of hydrodynamics in conservative lagrangian form are solved on a spherically symmetric 1D grid. Interface values are calculated via piecewise parabolic spatial reconstruction. The updated quantities are remapped to the original grid at the end of each step. This is the same as the piecewise parabolic method (PPM) with lagrangian remap (``PPMLR'') approach used by VH-1\footnote{http://wonka.physics.ncsu.edu/pub/VH-1/}. Energy losses are included via operator splitting using the exact integration scheme of \citet{Townsend:2009}. The solar abundance cooling curve from \citet{Wang:2014} is used in the low density limit. 
Solar abundances from \citet{Grevesse:2010} are adopted (mass fractions of $X=0.7381$, $Y=0.2485$, $Z=0.0134$).
Note that this gives a significantly reduced cooling rate compared to using the older \citet{Anders:1989} abundances \citep[as used, for example, by][]{Sutherland:1993}. Cooling is restricted at unresolved interfaces between hot diffuse gas and colder denser gas. At such interfaces the energy loss term is replaced with the minimum of the energy loss terms in the neighbouring cells. We do not include heating, or other physics such as thermal conduction and magnetic fields.

Analytical expressions for the time and radius of shell-formation, $t_{\rm SF}$ and $r_{\rm SF}$, depend on the adopted cooling rate, and therefore differ slightly between papers in the literature.  In this work we adopt the expressions given by \citet{Blondin:1998}:
\begin{equation}
\label{eq:tSF}
t_{\rm SF} \approx 2.9 \times 10^{4} E_{51}^{4/17} n_{\rm H}^{-9/17} \;\;{\rm yr},
\end{equation}
and
\begin{equation}
\label{eq:rSF}
r_{\rm SF} \approx 19.1 E_{51}^{5/17} n_{\rm H}^{-7/17} \;\;{\rm pc}.
\end{equation}
$E_{51}$ is the explosion energy in units of $10^{51}\,$erg, and the ambient intercloud density $\rho_{0} = n_{\rm H}m_{\rm H}/X = 1.355 n_{\rm H}m_{\rm H}$, $n_{\rm H}$ is the hydrogen number density in the ambient gas, where $X$ is the hydrogen mass fraction.

We evolve the SNR in an ambient medium of temperature $T_{\rm amb}=10^{4}$\,K and adopt a ratio of specific heats $\gamma=5/3$. A floor temperature is imposed for each grid cell which is set so that $T_{\rm floor}=T_{\rm amb}$ (we assume that the intercloud gas is kept photoionized by the environment). The SN ejecta is initialized as a region of uniform density with a linear velocity profile. We assume that the ejecta has a kinetic energy of $10^{51}$\,erg and a mass of $10\,\Msol$. This yields a maximum ejecta velocity of $4085\,\kmps$ and an initial radial momentum of $3.07\times10^{4}\,\Msol\,\kmps$ in all cases. The initial radius of the ejecta is assumed to be $0.02\,r_{\rm SF}$. The grid is set to a maximum radius of $5\,r_{\rm SF}$ and the calculations are evolved until $t=30\,t_{\rm SF}$. The calculations are run at high resolution to minimise the effects of numerical conduction at interfaces between cold and hot gas. We also want to accurately model the transition from the FE stage to the ST stage, and the process of shell formation. To these ends we use a grid containing 10,000 uniformly spaced cells. The ejecta is therefore mapped into the first 40 cells. The grid details are noted in Table~\ref{tab:grid}. 

\begin{table}
\centering
\caption[]{The details of the numerical grid. The maximum radial extent of the grid, $R_{\rm max}$, and the cell width, $dr$, are dependent on the ambient hydrogen number density.}
\label{tab:grid}
\begin{tabular}{lcr}
\hline
$n_{\rm H}$\,(cm$^{-3}$) & $R_{\rm max}$\,(pc) & $dr$\,(pc) \\
\hline
$10^{-2}$ &   636.1 & 0.0636\\	
1 &	95.50 &  0.00955 \\
$10^{2}$ & 14.34 & 0.00143 \\ 
\hline
\end{tabular}
\end{table}

We follow \citet{Kim:2015} and define ``hot'' gas as gas with $T > 2\times10^{4}$\,K, and shell gas as gas with $T \leq 2\times10^{4}$\,K and a radial velocity $v_{\rm r} > 1\,{\rm km\,s^{-1}}$. Since the time of shell formation in the simulations does not exactly agree with the analytical value from Eq.~\ref{eq:tSF}, we follow \citet{Kim:2015} and define the numerical time of shell formation, $t_{\rm SF}^{n}$, to be the time when the mass of hot gas, $M_{\rm hot}$, is a maximum. 

Several different definitions for the final radial momentum $p_{\rm final}$ exist in the literature. \citet{Martizzi:2015} defines this as ``the asymptotic radial momentum obtained after most of the thermal energy has been radiated away'', but do not note an explicit time that it is measured at. \cite{Walch:2015} measure theirs at $t=0.2\,$Myr ($t_{\rm SF}=8000$\,yr for their $\bar{n}=100\,{\rm cm^{-3}}$, so this is at $t/t_{\rm SF}=25$). \citet{Kim:2015} measure theirs at $t=10\,t_{\rm SF}^{n}$. The latter time is adopted in our uniform medium simulations. However, we note in Appendix~\ref{sec:tfloor} that $p_{\rm final}$ is weakly sensitive to the adopted value of $T_{\rm floor/amb}$ because the swept-up gas can contain non-negligible thermal energy that affects the late-time behaviour of the remnant. The mass-loaded simulations are more sensitive to the value of $T_{\rm floor/amb}$ (see Appendix~\ref{sec:tfloor}), and the radial momentum can increase significantly at late times due to a numerical effect associated with the imposed floor temperature (see Sec.~\ref{sec:withML} for further details). Therefore in our mass-loading simulations we instead measure $p_{\rm final}$ at $t=3\,t_{\rm SF}^{n}$. This timing avoids the numerical effects that occur later, but captures the majority of the post ST stage momentum boost.

\subsection{Details of the mass-loading}
The mass injection is assumed to take place from numerous sources (clouds) embedded within the remnant and is treated in the continuous limit (i.e. the intercloud spacing is small compared to the scale over which the properties of the remnant vary). We imagine that these clouds were initially located in the ambient medium, were dense enough to survive the initial passage of the forward shock over them, and are now being destroyed\footnote{Note that this is not equivalent to having a homogeneous background with some fluctuations - in such cases the remnant expansion speed simply reflects the local density that is being encountered and no mass injection occurs behind the shock.}. The mass injected from the clouds is assumed to have zero energy and momentum in the frame of the explosion. Mass injection occurs only within the volume of the remnant, between the initial ejecta radius and the remnant's forward shock. 

Several physical mechanisms may destroy the embedded clouds and/or mix their material into the background medium, including photoevaporation, hydrodynamic ablation, and conductively-driven thermal evaporation \citep[see][]{Pittard:2007}. Photoevaporation is unlikely to be the dominant cloud destruction process in this situation. The lifetime of clouds subject to thermal evaporation and hydrodynamic ablation depends on the properties of the cloud and flow (e.g., the cloud radius and density contrast; the flow density, velocity and temperature). Rather than attempting to deal with such complexity, we make the great simplification that the mass is injected at a constant volumetric rate. Following \citet{Dyson:2002} we define a fiducial mass loading rate
\begin{equation}
\label{eq:q0}
q_{0} = \frac{3v_{\rm SF}\rho_{0}}{r_{\rm SF}} \equiv \frac{6 \rho_{0}}{5 t_{\rm SF}} \,{\rm g\,cm^{-3}\,s^{-1}},
\end{equation}
where $v_{\rm SF}$ is the expansion velocity of the remnant at the time of shell formation. $q_{0}$ is the rate at which intercloud material is swept up divided by the remnant volume at the onset of shell formation. For $E_{51}=1.0$ and an ambient hydrogen number density $n_{\rm H}=0.01$, 1.0 and $100\,{\rm cm^{-3}}$, $q_{0} = 2.60\times10^{-39}$, $2.97\times10^{-36}$ and $3.40\times10^{-33}\,{\rm g\,cm^{-3}\,s^{-1}}$ respectively. The mass-loading is treated as a source term, $q=f_{\rm ML}q_{0}$, in the mass conservation equation, and is also treated via operator splitting. The relative dominance of mass-loading is set by the value of the variable $f_{\rm ML}$.

The assumption of a constant mass injection rate per unit volume means that the ratio of the mass injection rate to the rate that gas is swept up increases linearly with time \citep[see Eq.~11 in][]{Pittard:2003}. This causes the ratio of injected to swept-up {\em mass} to also increase linearly with time. As the ratio of masses is roughly comparable at $t=t_{\rm SF}^{n}$ if $f_{\rm ML}=1$, the injected mass becomes dominant for $t>t_{\rm SF}$ when $f_{\rm ML}=1$ (and becomes dominant for $t>10\,t_{\rm SF}$ when $f_{\rm ML}=0.1$).

\citet{Dyson:2002} also considered a Mach number dependence to their mass-loading rate for subsonic flow. However, we do not consider this dependence here for two reasons: first, the resulting evolution is little different from the constant rate case, and second the physical justification for this choice is no longer valid \citep[see][]{Pittard:2010}. In this work we also ignore mass-loading via conductively-driven thermal evaporation of embedded clouds. This process was studied by \citet{Pittard:2003}, but was found to be less effective in altering the remnant evolution because the mass-loading process shuts-off once the flow/remnant temperature drops below $10^{6}$\,K. 

We assume that there is a finite amount of mass in clouds, and define the parameter $\nu$ as the initial ratio of cloud to inter-cloud mass within a given region of the ISM. Larger values of $\nu$ mean that there is a larger ``reservoir'' of cloud mass available to be injected. A large value of $f_{\rm ML}$ and small value of $\nu$ is consistent with small, low density clouds that are rapidly destroyed after being engulfed by the remnant, so that most of the mass-loading occurs closely behind the forward shock. On the other hand, a large value of $\nu$ and smaller value of $f_{\rm ML}$ indicates that mass injection will occur throughout the remnant, and over long timescales, which is consistent with large and/or dense clouds that are long-lived. As clouds are destroyed and mass is injected the cloud mass that remains in each grid cell declines with time. Once this reaches zero no more mass can be injected into the remnant at this radius.

\begin{figure*}
\includegraphics[width=17.5cm]{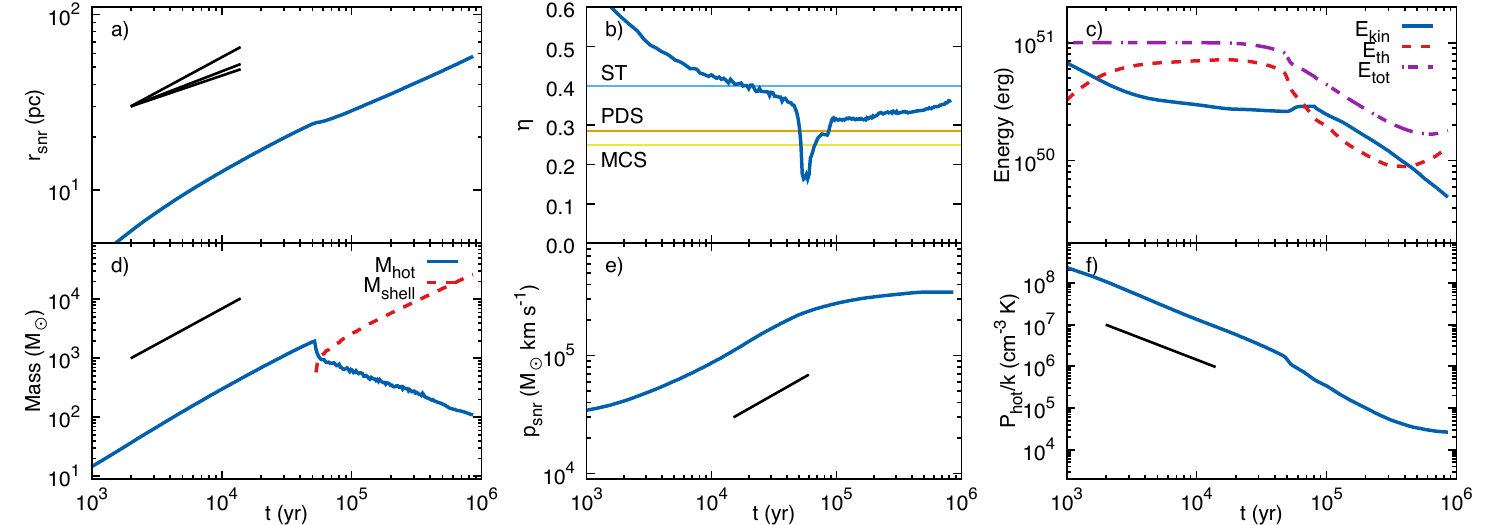}
\caption{The time evolution of model nH1fML0 ($n_{\rm H} = 1\,{\rm cm^{-3}}$, $f_{\rm ML}=0$; no mass-loading). The panels show: a) the radius; b) the deceleration parameter $\eta \equiv v_{\rm snr}t/r_{\rm snr}$; c) the total, thermal and kinetic energies; d) the mass of interior ``hot'' gas and the shell mass; e) the total radial momentum; f) the pressure of interior ``hot'' gas. The initial ejecta radius for this model is 0.382\,pc. The analytical scalings of radius with time and the values of the deceleration parameter for the ST, PDS and MCS stages are shown in panels a) and b). The analytical scalings of the remnant mass, radial momentum and pressure with time during the ST stage are shown in panels d)-f).}
\label{fig:nH1fML0_stats}
\end{figure*}

\begin{figure*}
\includegraphics[width=17.5cm]{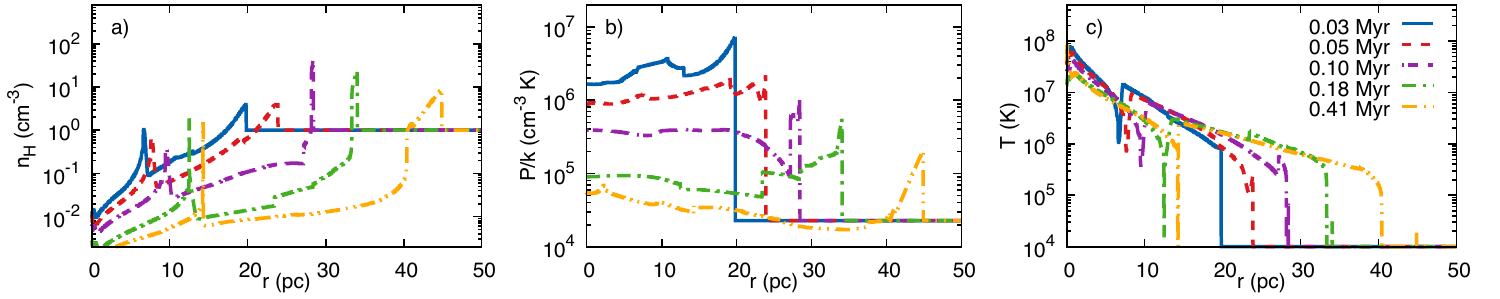}
\caption{Radial profiles of model nH1fML0 ($n_{\rm H} = 1\,{\rm cm^{-3}}$, $f_{\rm ML}=0$; no mass-loading) at selected times. The panels show: a) number density; b) pressure; c) temperature.}
\label{fig:nH1fML0_profiles}
\end{figure*}

\section{Results}
\label{sec:results} 

We examine first how the remnant evolves when there is no mass-loading. Then we consider the effect of mass-loading. Finally, we consider the effect of multiple explosions. We adopt a naming convention for our models such that a model with name nH$A$fML$B$nu$C$ has an intercloud ambient hydrogen density of $A\,{\rm cm^{-3}}$, a mass-loading rate of $f_{\rm ML} = B$ and a ratio of cloud to intercloud mass $\nu = C$. Models with no mass-loading have truncated names (e.g., nH$A$fML0).

\subsection{Single SN in a uniform medium (no mass-loading)}
\label{sec:noML}
We first present results for model nH1fML0 ($n_{\rm H}=1\,{\rm cm^{-3}}$, $f_{\rm ML}=0$). Fig.~\ref{fig:nH1fML0_stats} shows the time evolution of a) the radius, $r_{\rm snr}$; b) the deceleration parameter, $\eta \equiv v_{\rm snr}t/r_{\rm snr}$, where $v_{\rm snr}$ is calculated from the time taken for $r_{\rm snr}$ to change by $\pm$1 per cent ($\eta$ is not smoothed); c) the total, $E_{\rm tot}$, thermal, $E_{\rm th}$, and kinetic, $E_{\rm kin}$, energies; d) the mass of interior ``hot'' gas, $M_{\rm hot}$, and shell gas, $M_{\rm shell}$; e) the total radial momentum, $p_{\rm snr}$; f) the mean pressure of interior ``hot'' gas, $P_{\rm hot}=\int_{\rm hot} PdV/\int_{\rm hot} dV$. Fig.~\ref{fig:nH1fML0_profiles} shows the profiles of various fluid variables at selected times.

The remnant is initially ejecta dominated. A reverse shock forms which decelerates the ejecta, while the ejecta drives a forward shock into the ambient medium. The reverse shock initially moves outwards in the frame of the explosion, but after $\approx 3000$\,yr it starts to move back towards the origin, arriving at $\approx 7000$\,yr. At this point all of the ejecta is thermalized. A reflected shock moves back out and arrives at the contact discontinuity between the ejecta and the swept-up ambient medium at $t \approx$13,000\,yr. At $t=0.03$\,Myr (the first snapshot in Fig.~\ref{fig:nH1fML0_profiles}) this shock is roughly half-way between the contact discontinuity and the forward shock, and it continues to reverberate throughout the remnant thereafter as the ``echoes of thunder'' noted by \citet{Cioffi:1988}. By $t=0.02$\,Myr, the remnant has swept up $700\,\Msol$ of material and has transitioned into a quasi-ST stage (in the ST stage, $\eta=0.4$, $E_{\rm th}/E_{\rm tot} = 0.717$, and $E_{\rm kin}/E_{\rm tot} = 0.283$).  

The profiles in Fig.~\ref{fig:nH1fML0_profiles} show a sharp density jump at the contact discontinuity that persists as the remnant evolves. This forms from the initial conditions and is partially numerical in origin: in a real remnant, and in 3D simulations, it is subject to Rayleigh-Taylor (RT) instabilities and soon smooths out. However, this is prevented in our 1D simulations. Similarly, we expect the cold shell that forms at the end of the ST stage to be subject to RT instabilities and so to be broader and less dense in 3D simulations. This may reduce the radiative cooling near the shell in the 3D simulations, but it is not expected to lead to significant differences in the global behaviour of the remnant given the significant broadening of the shell that occurs as the Mach number of the forward shock decreases over time.

During the ST stage the momentum of the remnant increases as $t^{3/5}$, which is indeed what we see in Fig.~\ref{fig:nH1fML0_stats}. The radiative stage for model nH1fML0 begins at $t_{\rm SF}^{n} = 0.0525$\,Myr, when the total mass of hot gas peaks at $2000\,\Msol$. The reflected shock has not reached the forward shock at this time. A thin dense shell rapidly forms behind the forward shell and a complicated series of shocks, rarefaction waves and contact discontinuities also form \citep[see, e.g.,][for details]{Falle:1975}. During shell formation the gas loses its forward momentum as its thermal pressure support against the ram pressure of the oncoming ambient gas disappears. Shortly afterwards $\eta$ reaches a minimum value, but then increases as the pressure of the hot interior gas drives the shell forwards once more. $\eta$ eventually settles near the value expected in the PDS stage (where $\eta=2/7=0.286$). However, since $M_{\rm hot}$ continues to decrease, a ``cooling-modified'' PDS stage actually occurs. Once formed, the mass of the shell continues to increase.  At very late times the thermal energy stored in the swept-up gas (which is at temperature $T_{\rm amb}=10^{4}$\,K in these calculations) begins to become significant and $E_{\rm th}$ and $E_{\rm tot}$ both rise. This also affects the total radial momentum at late times (see Appendix~\ref{sec:tfloor}).

The behaviour of the remnant agrees with previously published work \citep[e.g.,][]{Kim:2015}. In their equivalent calculation, \citet{Kim:2015} find that $t_{\rm SF}^{n} = 0.0419$\,Myr. The difference between our result and theirs is likely caused by their use of fits to the cooling function of \citet{Sutherland:1993}, which is stronger than the lower metallicity solar abundance cooling curve adopted here. We find a total radial momentum at $t_{\rm SF}^{n}$ of $p_{\rm SF}^{n} = 2.26\times10^{5}\,\Msol\,{\rm km s^{-1}}$, and a final radial momentum at $t=10\,t_{\rm SF}^{n}$ of $p_{\rm final} = 3.46\times10^{5}\,\Msol\,{\rm km s^{-1}}$. These values are 10 and 15 per cent higher, respectively, than found by \citet{Kim:2015}. Some of this difference will be due to the higher value of $T_{\rm amb}=T_{\rm floor}$ that we use in our simulations compared to \citet{Kim:2015} (see Appendix~\ref{sec:tfloor}).

Remnants evolve more (less) rapidly at higher (lower) ambient density. The longer time to shell formation at lower densities means that the remnant can do more PdV work on the swept-up gas, resulting in higher values of $p_{\rm SF}^{n}$ and $p_{\rm final}$. In Table~\ref{tab:noML_results} we note the values of certain physical quantities at the time of numerical shell formation for different values of $n_{\rm H}$. A fit to the values of $p_{\rm SF}^{n}$ gives
\begin{equation}
\label{eq:pSF_fit}
p_{\rm SF}^{n} = (2.21\pm0.05) \times 10^{5} n_{\rm H}^{-0.140\pm0.006} \,\Msol\,{\rm km\,s^{-1}}.
\end{equation}
The fit is in good agreement with Eq.~25 in \citet{Kim:2015}. 

The final momentum attained, $p_{\rm final} = 6.03$, 3.46 and $1.59\times 10^{5}\,\Msol\,{\rm km\,s^{-1}}$ when $n_{\rm H}=0.01$, 1 and $100\,{\rm cm^{-3}}$, respectively. A fit to these values gives
\begin{equation}
\label{eq:pfinal_fit}
p_{\rm final} = (3.3\pm0.2) \times 10^{5} n_{\rm H}^{-0.13\pm0.01} \,\Msol\,{\rm km\,s^{-1}}.
\end{equation}
This is within $2\sigma$ of Eq.~29 in \citet{Kim:2015}. Since the momentum of the ejecta, $p_{\rm ejecta} = 3.07\times10^{4} \,\Msol\,{\rm km\,s^{-1}}$, this implies that there is no hot gas and PdV work when $n_{\rm H} \gtsimm 5\times10^{7} \,{\rm cm^{-3}}$.

\begin{table}
\centering
\caption[]{Physical quantities at the time of numerical shell formation in a uniform density ambient medium with no mass-loading. The ambient mass density $\rho = 1.355\, n_{\rm H}m_{\rm H}$.}
\label{tab:noML_results}
\begin{tabular}{ccccc}
\hline
Model & $n_{\rm H}$ & $t_{\rm SF}^{n}$ & $M_{\rm hot}$ & $p_{\rm SF}^{n}/10^{5}$\\
      & (cm$^{-3}$) & (kyr) & ($\Msol$) & ($\Msol\,{\rm km\,s^{-1}}$)\\
\hline
nH0.01fML0 & 0.01& 698 & 6460 & 4.2\\	
nH1fML0    & 1.0      &	52.5 & 2000 & 2.26\\
nH1e2fML0  & $10^{2}$ & 3.72 & 520 & 1.11\\ 
\hline
\end{tabular}
\end{table}

\begin{figure*}
\includegraphics[width=17.5cm]{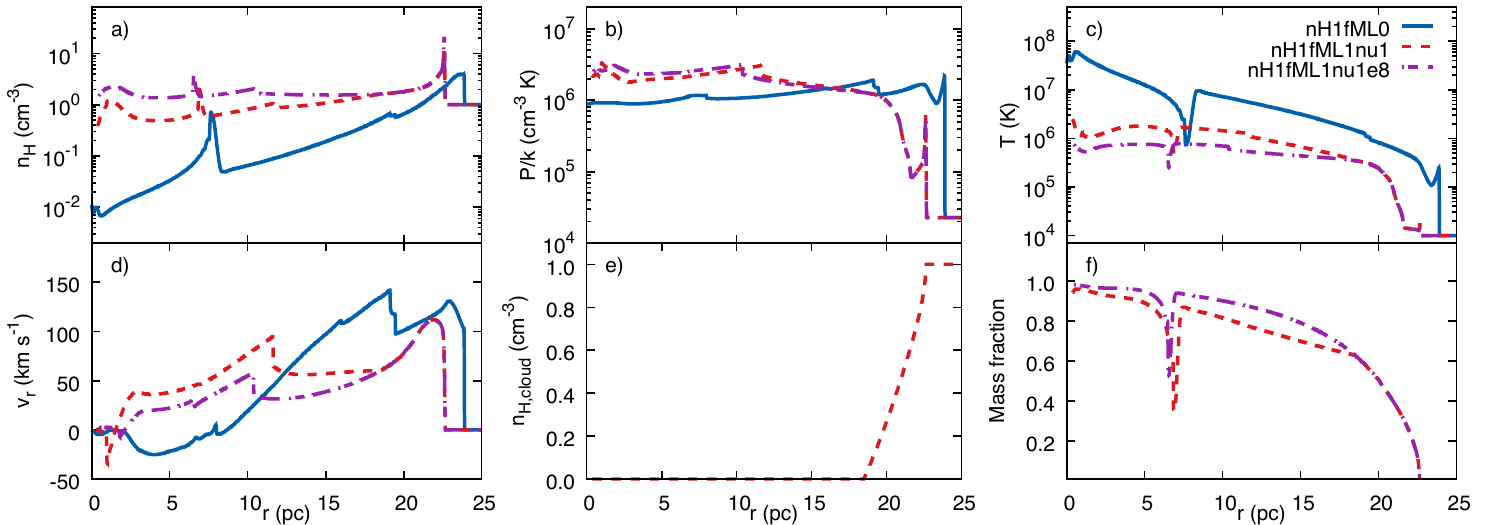}
\caption{Radial profiles at $t=0.05$\,Myr of models nH1fML0 ($n_{\rm H} = 1\,{\rm cm^{-3}}$, $f_{\rm ML}=0$; no mass-loading), nH1fML1nu1 ($n_{\rm H} = 1\,{\rm cm^{-3}}$, $f_{\rm ML}=1$, $\nu=1$; mass-loading with a finite reservoir of cloud mass), and nH1fML1nu1e8 ($n_{\rm H} = 1\,{\rm cm^{-3}}$, $f_{\rm ML}=1$, $\nu=10^{8}$; mass-loading with effectively an infinite reservoir of cloud mass). The panels show: a) the number density; b) the pressure; c) the temperature; d) the radial velocity; e) the smeared-out density of cloud material; and f) the fraction of injected mass.}
\label{fig:nH1fML0and1_profiles}
\end{figure*}

\begin{figure*}
\includegraphics[width=17.5cm]{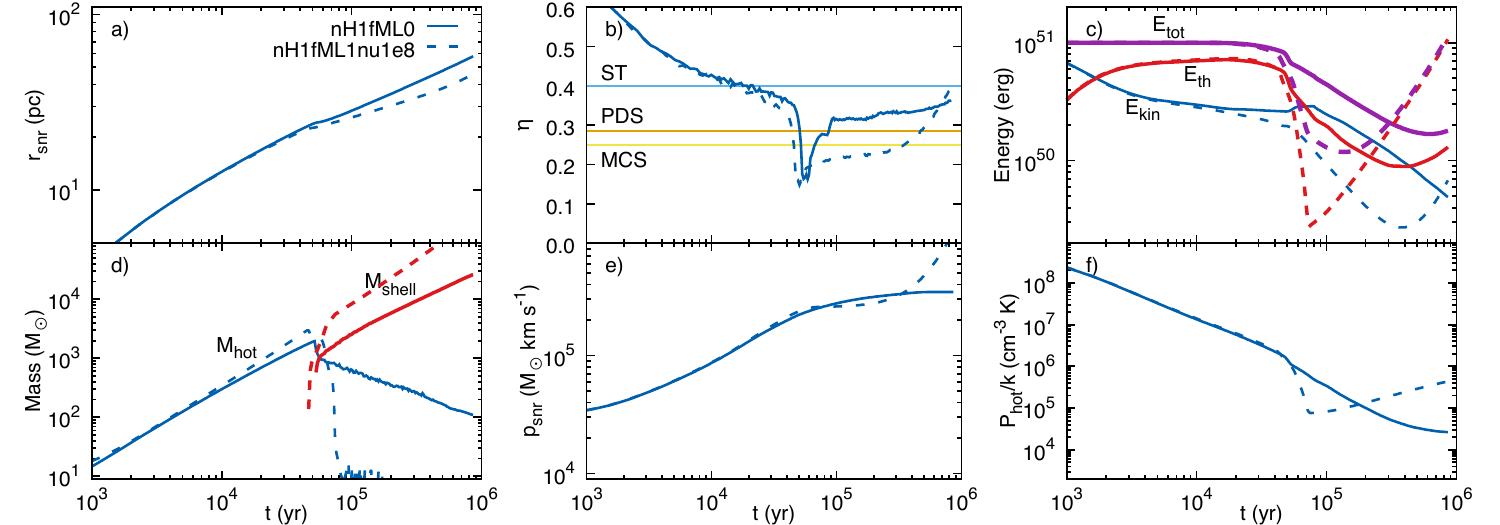}
\caption{The time evolution of model nH1fML1nu1e8 (dashed lines, $n_{\rm H} = 1\,{\rm cm^{-3}}$, $f_{\rm ML}=1$, $\nu=10^{8}$; mass-loading). The panels show: a) the radius; b) the deceleration parameter; c) the total, thermal and kinetic energies; d) the mass of interior ``hot'' gas and shell gas; e) the total radial momentum; f) the pressure of interior ``hot'' gas. Model nH1fML0 (solid lines; no mass-loading) is also shown for comparison. The analytical values of the deceleration parameter for the ST, PDS and MCS stages are shown in panel b).}
\label{fig:nH1variousfML_stats}
\end{figure*}

\begin{figure*}
\includegraphics[width=17.5cm]{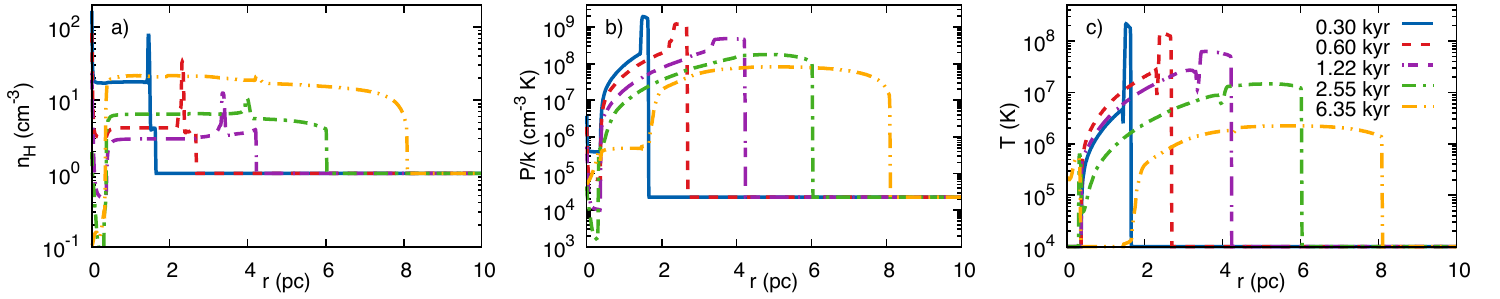}
\caption{Radial profiles of model nH1fML100nu1e8 ($n_{\rm H} = 1\,{\rm cm^{-3}}$, $f_{\rm ML}=100$, $\nu=10^{8}$; mass-loading) at selected times. The panels show: a) number density; b) pressure; c) temperature.}
\label{fig:nH1fML100_profiles}
\end{figure*}

\begin{figure*}
\includegraphics[width=17.5cm]{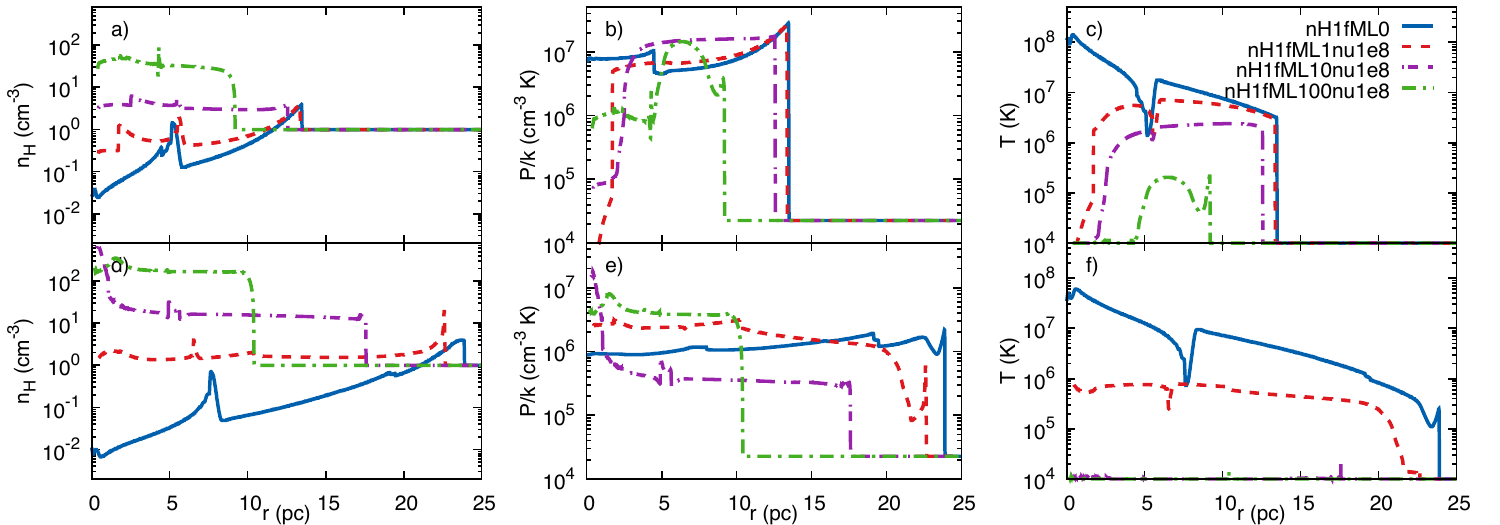}
\caption{Radial profiles of models nH1fML0 ($n_{\rm H} = 1\,{\rm cm^{-3}}$, $f_{\rm ML}=1$; no mass-loading), nH1fML1nu1e8, nH1fML10nu1e8, and nH1fML100nu1e8. The mass-loading models all have an infinite reservoir of cloud mass but inject mass into the remnant at different rates. The top row shows snapshots at $t=$11,500\,yr, while the bottom row shows snapshots at $t=$50,500\,yr.}
\label{fig:nH1variousfML_profiles}
\end{figure*}

\begin{figure*}
\includegraphics[width=17.5cm]{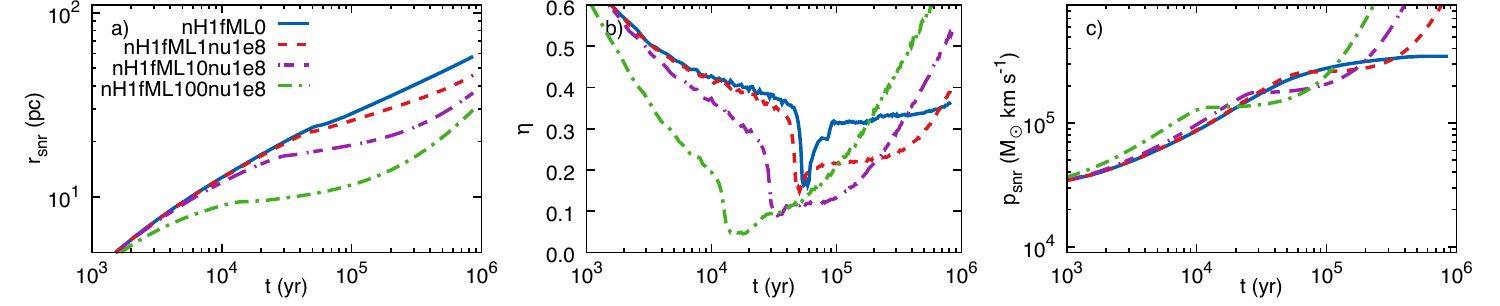}
\caption{The time evolution of models nH1fML0 (no mass-loading), nH1fML1nu1e8, nH1fML10nu1e8, and nH1fML100nu1e8. The panels show: a) the radius; b) the deceleration parameter; c) the total radial momentum.}
\label{fig:nH1variousfMLnu1e8_stats}
\end{figure*}

\subsection{Single SN in an inhomogeneous medium (with mass-loading)}
\label{sec:withML}

We now consider the evolution of remnants expanding into an inhomogeneous medium, where there is mass injection from clouds engulfed by, and embedded within, the remnant. Since the ratio of the rate of mass injection to the rate at which intercloud mass is swept up increases linearly with time, mass-loading becomes increasingly dominant and/or important with time.

We first consider the evolution of remnants expanding into an intercloud medium of $n_{\rm H} = 1\,{\rm cm^{-3}}$, for the case where the mass injection rate is $f_{\rm ML}=1$ (i.e. at the time of shell formation, mass is being injected into the remnant at roughly the same rate as it is being swept up). Fig.~\ref{fig:nH1fML0and1_profiles} shows radial profiles at $t=0.05$\,Myr for models nH1fML1nu1 (mass-loading with a finite reservoir of cloud mass) and nH1fML1nu1e8 (mass-loading with effectively an infinite reservoir of cloud mass). Model nH1fML0 (no mass-loading) is also shown for comparison. Since model nH1fML1nu1 has a ratio of cloud to intercloud mass of 1, one could imagine that the clouds are on average $9\times$ denser but occupy only 10 per cent of the volume of the ISM (any other suitable combination could also be imagined, such as clouds that are on average $99\times$ denser with a volume filling fraction of 1 per cent).

Mass-loading into the interior of the remnant has several noticeable effects. Early on, the mass injection that occurs within the unthermalized ejecta (but outside the original radius of the ejecta) drag heats it to high temperatures, and reduces its pre-shock velocity. Mass injection into the shocked region between the reverse shock and the forward shock also slows this flow, and in general reduces its temperature as there are now more particles to share the thermal energy between\footnote{More generally, mass-loading of an adiabatic flow tends to drive the flow to a Mach number $\cal{M}$ of unity \citep{Hartquist:1986}.}. If the mass loading that is taking place is vigorous enough, it can cause the density of the unthermalised ejecta to start {\em increasing}, and it can alter the behaviour and position of the reverse shock. In model nH1fML1nu1e8, the unthermalised ejecta starts to {\em increase} its density by $t=4$000\,yr. By $t=1$5,250\,yr the reverse shock has reached the centre of the remnant (i.e. taking more than twice as long to thermalize all of the ejecta compared to the model with no mass loading) and once again expands outwards.

In the models shown in Fig.~\ref{fig:nH1fML0and1_profiles}, by $t=0.05$\,Myr mass-loading has visibly increased the remnant density. This is most obvious near the centre of the remnant where the density in the model without mass-loading (model nH1fML0) is lowest. The temperature within the mass-loaded remnants has also become more uniform, and is currently at $T\sim10^{6}$\,K. The increase in density and decrease in temperature increases the cooling rate of the gas, and we see that shell formation occurs earlier. Thus mass-loading speeds up the remnant evolution and causes the ST stage to end sooner.

Clouds nearest to the explosion are engulfed first as the remnant expands, with clouds at greater distances becoming embedded at later times as the remnant expands further. Thus, for the case of a finite mass reservoir with a uniform cloud distribution and where the mass injection rate per unit volume is also uniform, the available mass to inject will first run out at small radii, since the clouds there have been embedded in the remnant the longest. Fig.~\ref{fig:nH1fML0and1_profiles}e) shows that all of the available cloud mass out to $r=18$\,pc has been loaded into model nH1fML1nu1 by this time. Thus mass-loading is currently taking place only between $18 \ltsimm r/{\rm pc} \ltsimm 22.7$\footnote{This is somewhat similar to the temperature-dependent mass-loading used in \citet{Pittard:2003}, where only regions of the remnant with $T \gtsimm 10^{6}$\,K are still being mass-loaded.}. Therefore the profiles from models nH1fML1nu1 and nH1fML1nu1e8 are identical within this region, but differ at $r<18$\,pc due to the remnant in model nH1fML1nu1 being ``starved'' of available clouds for continued mass-loading.

Fig.~\ref{fig:nH1variousfML_stats} shows the time evolution of various statistics for model nH1fML1nu1e8, with model nH1fML0 shown for comparison. Mass-loading initially increases the mass of hot gas within the remnant, and increases the thermal energy fraction at the expense of the kinetic energy fraction. Mass-loading also reduces the expansion rate of the remnant and increases its total radial momentum in this early period. However, the increased interior densities cause the remnant to leave the ST stage sooner. All the hot gas within the remnant disappears soon after shell formation so that by $t=10^{5}$\,yr there is essentially none remaining. This differs markedly from the behaviour of models without mass-loading, where a substantial amount of hot gas remains within the remnant for long times (due to the very long cooling time of the low density gas near the centre of the remnant). This rapid loss of hot gas means that the mass-loaded remnant in model nH1fML1nu1e8 has a shorter and less significant PDS stage. At the beginning of the MCS stage, at $t=10^{5}$\,yr, the deceleration parameter, $\eta\approx0.22$. This is slightly lower than the value expected (in the MCS stage, $\eta=1/4$), and reflects the on-going mass-loading that is taking place. Cold gas now exists throughout most of the remnant, but the majority of the mass of the cold gas is near the shock front. This is not the case in simulations with more vigorous mass-loading (e.g., nH1fML10nu1e8) where the cold mass is distributed at roughly uniform density throughout the remnant (see Fig.~\ref{fig:nH1variousfML_profiles}). The cold ``shell'' mass continues to climb as further mass is continually injected into, and swept up by, the remnant (see Fig.~\ref{fig:nH1variousfML_stats}d).

The total mass within the remnant in model nH1fML1nu1e8 reaches $2.5\times10^{5}\,\Msol$ at $t=0.9\,$Myr. This is $10\times$ higher than the remnant without mass-loading. The swept-up ambient mass is $1.35\times10^{4}\,\Msol$, which represents only 5 per cent of the total remnant mass. As previously noted, since the ratio of the mass injection rate to the rate at which intercloud mass is swept up increases linearly with time, the injected mass dominates the swept up mass when $t > t_{\rm SF}$ if $f_{\rm ML}=1.0$. Most of the injected mass at the end of the simulation ($t=0.9$\,Myr) has therefore been recently injected.

Fig.~\ref{fig:nH1variousfML_stats}c) shows that the thermal energy reaches a minimum value of $\approx 3\times10^{49}$\,erg at $t\approx70,000$\,yr, and thereafter rises significantly. The latter behaviour is numerical in origin and is caused by the imposed floor temperature of $T_{\rm floor}=10^{4}$\,K. Although mass is injected with zero energy, if mass is injected into a cell where the initial cell temperature $T = T_{\rm floor}$, the new cell temperature will be below $T_{\rm floor}$, and will be reset to $T = T_{\rm floor}$ at the end of the numerical step. For cells where this occurs (which only happens after the end of the mass-loaded-modified ST stage), mass is effectively being injected with a temperature of $T_{\rm floor}$, and with the thermal energy associated with this. At $T = 10^{4}$\,K, $1\,\Msol$ of material has a thermal energy of $4\times10^{45}$\,erg. Injecting $2.5\times10^{5}\,\Msol$ of material at $T_{\rm floor} = 10^{4}$\,K supplies $10^{51}$\,erg of thermal energy, which is what we see in Fig.~\ref{fig:nH1variousfML_stats}c) and d). The extra energy powers a late-time expansion of the remnant which drives a further increase in the radial momentum. $\eta$ rises towards and then past the PDS value. Since $\eta$ is increasing with time the remnant's deceleration is slowing. At $t=0.9$\,Myr its expansion velocity has plateaued at $v=21\,\kmps$. This behaviour arises because the pressure within the remnant is now increasing with time (mass is continually injected, increasing the density, while the gas is maintained at an effective floor temperature of $10^{4}$\,K). Setting a lower value for $T_{\rm floor}$ reduces this effect (see Appendix~\ref{sec:tfloor}).

\subsubsection{Comparison of different mass-loading rates with an infinite reservoir of cloud mass}
\label{sec:variousfMLinfiniteReservoir}
We now wish to compare the evolution of remnants that experience different rates of mass-loading from an infinite reservoir of cloud mass. The first thing to note is that when $f_{\rm ML} >> 1$, mass injection is so rapid that the remnant evolution is strongly modified at the earliest stages, even before the ejecta has been completely thermalized. This is illustrated in Fig.~\ref{fig:nH1fML100_profiles} which shows various profiles from a simulation with $n_{\rm H}=1.0\,{\rm cm^{-3}}$ and $f_{\rm ML}=10^{2}$. In this case we see that the rapid injection of mass drag-heats the unthermalized ejecta to a temperature in excess of $10^{7}\,$K {\em before} it encounters the reverse shock, while the density of the unthermalized ejecta starts to {\em increase} with time after $t=1200$\,yr. The peak density of the shocked ejecta declines relative to its surroundings as mass-loading continues, effectively burying it under the general increase in density. The pre-shock ejecta, the shocked ejecta, and the shocked ambient gas all drop in temperature as time proceeds so that by $t=3200$\,yr all of the gas in the remnant is below a temperature of $10^{7}$\,K. The velocity of the mass-loaded ejecta also drops rapidly with time and in this simulation, the reverse shock does not make it back to the centre of the remnant. By $t=3600$\,yr some of the frictionally-heated ejecta has cooled back to $10^{4}$\,K.

Fig.~\ref{fig:nH1variousfML_profiles} shows the later evolution of this simulation, with snapshots shown at $t=11$,500\,yr (top row) and at $t=50$,500\,yr (bottom row). At $t=11$,500\,yr, mass-loading has raised the central density of the remnant to a level that is $30\times$ that of the ambient intercloud density. Panel c) in Fig.~\ref{fig:nH1variousfML_profiles} shows that by $t=11$,500\,yr, all of the frictionally-heated ejecta at $r<5$\,pc is now at $10^{4}$\,K in the $f_{\rm ML}=10^{2}$ simulation. The gas above $10^{4}$\,K that is located at $r>5$\,pc is the gas that was actually heated by either the reverse or the forward shock. This gas is radiating strongly and cools to $10^{4}$\,K within the next 500\,yr.

The nH1fML10nu1e8 simulation shares many characteristics with the n1fML100nu1e8 simulation just described. Its unthermalised ejecta starts increasing in density by $t=2$200\,yr, and attains a near uniform level by $t=11$,500\,yr. The reverse shock also does not make it back to the centre (it is at $r=2.5$\,pc in the top panels of Fig.~\ref{fig:nH1variousfML_profiles}; in model nH1fML1nu1e8 the reverse shock is visible in Fig.~\ref{fig:nH1variousfML_profiles}a) at $r=1.7$\,pc). 

By $t=50$,500\,yr (the lower panels in Fig.~\ref{fig:nH1variousfML_profiles}) the entire remnant consists of cold gas in the simulations with $f_{\rm ML} \geq 10$. The model with no mass-loading has a hot interior but is about to form its cold, outer shell, while the model with $f_{\rm ML}=1$ has just formed its shell (the gas at radii between $r=21.4-22.7$\,pc does not immediately cool to $T=10^{4}$\,K, but seems to be maintained at $\approx$13,500\,K by frictional heating from the ongoing mass injection).

Fig.~\ref{fig:nH1variousfML_profiles} and~\ref{fig:nH1variousfMLnu1e8_stats} show just how much the early expansion of the remnant is hindered by heavy mass-loading. As before, if mass-loading is not limited in any way, at late times the remnant stops decelerating (and reaches a constant expansion speed of $20\,\kmps$), and the radial momentum significantly increases. As previously noted this is caused by setting a floor temperature of $10^{4}$\,K. At early times, the injection of mass causes the radial momentum to initially increase faster, but then to plateau at an earlier time (due to the hot gas disappearing) and at a lower level. The model with the most vigourous mass-loading, nH1fML100nu1e8, plateaus at a radial momentum of $1.35\times10^{5} \,\Msol\,{\rm km\,s^{-1}}$, which is nearly $3\times$ lower than the model without mass-loading. 

\begin{figure*}
\includegraphics[width=17.5cm]{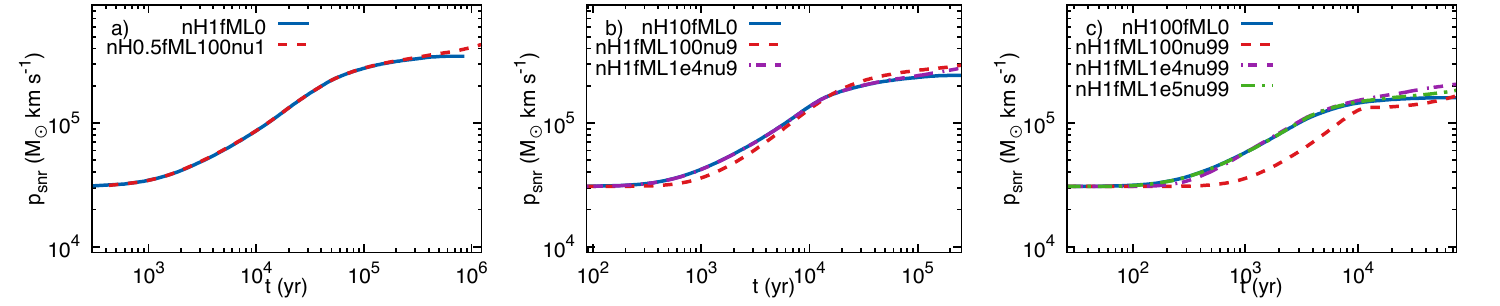}
\caption{The time evolution of the total radial momentum for models with an average ambient hydrogen number density of: a) $\bar{n}_{\rm H}=1\,{\rm cm^{-3}}$; b) $\bar{n}_{\rm H}=10\,{\rm cm^{-3}}$; c) $\bar{n}_{\rm H}=10^{2}\,{\rm cm^{-3}}$. In each panel models with the same average (cloud plus intercloud) density are compared.}
\label{fig:massLoadMimicHighernH_mtm}
\end{figure*}
 
\subsubsection{Equivalence of mass-loaded remnants to remnants expanding in uniform media}
\label{sec:equivalence}
Fig.~\ref{fig:nH1variousfMLnu1e8_stats} shows that mass-loading alters the evolution of the momentum of the remnant, and the final momentum that is obtained. An obvious question is whether remnants that experience very rapid mass-loading from a finite mass reservoir evolve similar to remnants expanding into a uniform medium of the same average (cloud plus intercloud) density. If the mass-loading is very rapid but limited it should be confined to a small region behind the forward shock and occur at a rate that is equivalent to the rate at which mass is swept up by the shock in the equivalent density uniform medium. Fig.~\ref{fig:massLoadMimicHighernH_mtm} shows that such simulations are indeed basically equivalent, provided that mass-loading occurs quickly enough. Fig.~\ref{fig:massLoadMimicHighernH_mtm}a) shows that a simulation with an intercloud ambient density $n_{\rm H}=0.5\,{\rm cm^{-3}}$, a cloud to intercloud mass ratio $\nu=1$ (giving a mean ambient density $\bar{n}_{\rm H}=1\,{\rm cm^{-3}}$), and a rapid rate of mass injection ($f_{\rm ML}=100$) does indeed evolve almost identically to a remnant expanding into a uniform medium with $n_{\rm H}=1\,{\rm cm^{-3}}$. The only significant difference occurs at late times due to the rise in thermal energy caused by injecting gas at the floor temperature (see previous section). 

Fig.~\ref{fig:massLoadMimicHighernH_mtm}b) and~c) show that this equivalence extends to environments that are increasingly inhomogeneous, though even faster rates of injection are needed. Comparing model nH10fML0 and nH1fML100nu9 in Fig.~\ref{fig:massLoadMimicHighernH_mtm}b) we see that the momentum in the mass-loaded $f_{\rm ML}=100$ model starts to rise later, but then eventually exceeds that of the uniform model. Setting $f_{\rm ML}=10^{4}$ causes the momentum to rise at the same time as the uniform model and for a nearly identical final momentum to be obtained. Fig.~\ref{fig:massLoadMimicHighernH_mtm}c) shows that $f_{\rm ML}=10^{5}$ is needed to obtain identical early-time behaviour when $\nu = 99$.
Thus, in the extreme case that limited mass injection occurs very rapidly, the remnants behave like they are expanding into a medium of the equivalent average density. The extreme values of $f_{\rm ML}$ that are required for this equivalence are needed to ensure that the density behind the forward shock rises rapidly to its postshock equivalent in the uniform medium case (e.g., when the intercloud medium has $n_{\rm H}=1\,{\rm cm^{-3}}$, the postshock density (for a strong shock and $\gamma=5/3$) is $n_{\rm H}=4\,{\rm cm^{-3}}$, but this needs to be rapidly increased by mass-loading to $n_{\rm H}=400\,{\rm cm^{-3}}$ if the mass-loaded solution is to be equivalent to a remnant expanding into a uniform medium of density $n_{\rm H}=100\,{\rm cm^{-3}}$).

In general we do not expect rates of mass injection as extreme as $f_{\rm ML}=10^{4}$. When mass injection occurs at a more measured pace, and throughout the remnant rather than in a narrow region behind the forward shock, it causes the remnant to evolve differently, sometimes very differently, as Figs.~\ref{fig:nH1fML0and1_profiles}-\ref{fig:nH1variousfMLnu1e8_stats} show. Thus, in general, we find that the mass-loading simulations do not follow the scaling with mean density found by \citet{Kim:2015}, if the gas is able to effectively mix with the remnant gas. This is a key difference between their work and this work.

\subsubsection{Fits to $t_{\rm SF}^{n}$, $r_{\rm SF}^{n}$, $p_{\rm SF}^{n}$ and $p_{\rm final}$}
\label{sec:radial_mtm}

\begin{figure*}
\includegraphics[width=17.5cm]{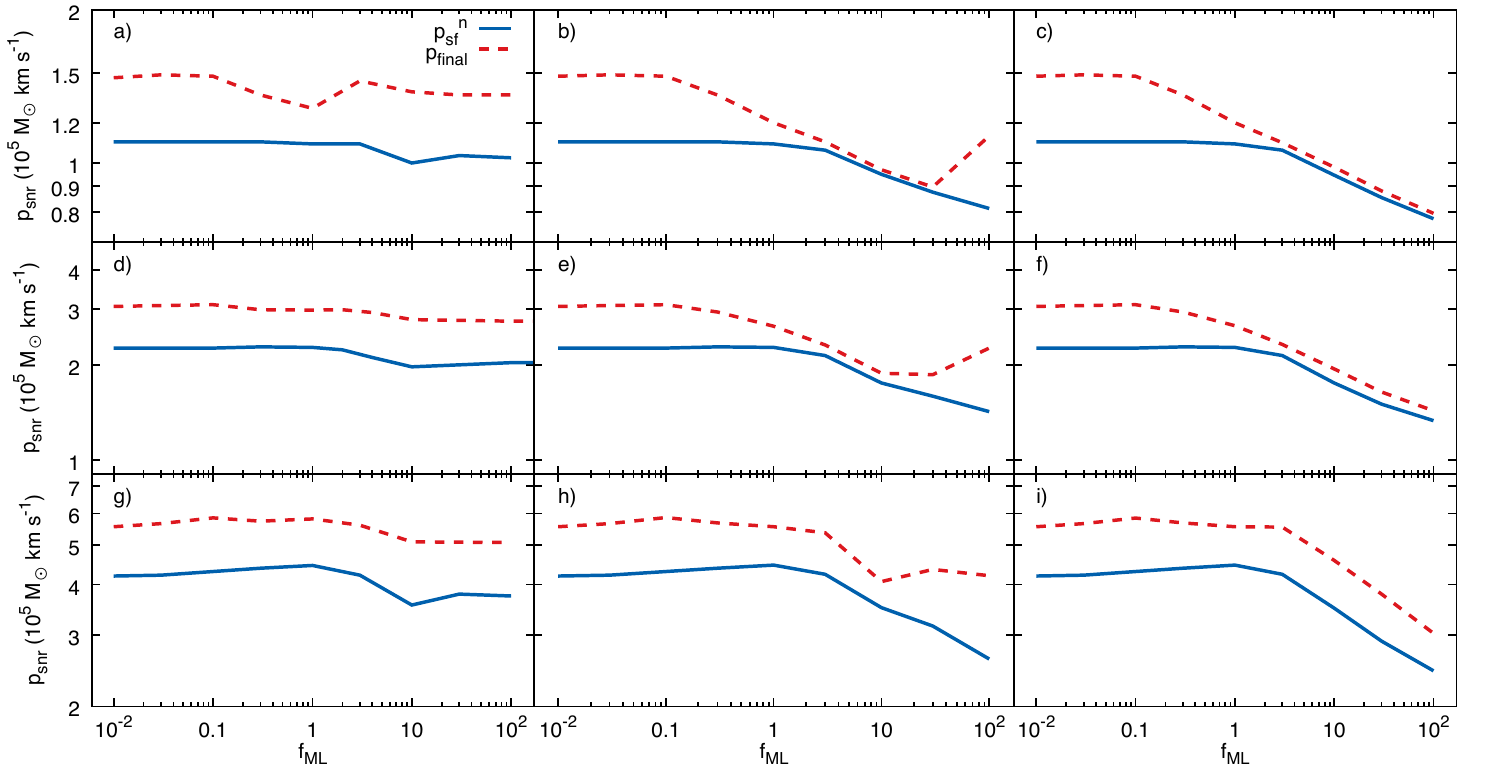}
\caption{The total radial momentum at $t=t_{\rm SF}^{n}$ ($p_{\rm SF}^{n}$) and at $t=3\,t_{\rm SF}^{n}$ ($p_{\rm final}$) as a function of $\nu$, $f_{\rm ML}$ and $n_{\rm H}$. The top, middle and bottom rows show results for $n_{\rm H}=100\,{\rm cm^{-3}}$, $n_{\rm H}=1\,{\rm cm^{-3}}$ and $n_{\rm H}=0.01\,{\rm cm^{-3}}$ respectively. The left, middle and right columns show results for $\nu=1$, $\nu=10$ and $\nu=10^{8}$ respectively.}
\label{fig:varfML_mtm}
\end{figure*}

Fig.~\ref{fig:varfML_mtm} shows the radial momentum $p_{\rm SF}^{n}$ (at $t=t_{\rm SF}^{n}$) and $p_{\rm final}$ (measured at $t=3\,t_{\rm SF}^{n}$ - i.e. before the floor temperature becomes a numerical issue in any of the simulations). The remnant is increasingly mass-loaded as $f_{\rm ML}$ rises, has a shorter ST stage, and achieves a lower value of $p_{\rm SF}^{n}$ and $p_{\rm final}$. However, at low values of $\nu$, the finite cloud mass can curtail further mass-loading: this leads to $p_{\rm SF}^{n}$ and $p_{\rm final}$ becoming independent of $f_{\rm ML}$ when $\nu=1$ and $f_{\rm ML}\gtsimm 10$. The $\nu=10$ simulations show behaviour between that of the $\nu=1$ and $\nu=10^{8}$ models. 

When there is an infinite amount of cloud mass that can be added into the remnant, the ratio of $p_{\rm final}/p_{\rm SF}^{n}$ decreases as $f_{\rm ML}$ increases. At high $n_{\rm H}$ and $f_{\rm ML}$, $p_{\rm final}/p_{\rm SF}^{n}$ becomes close to unity (i.e. there is very little ``boost'' to the radial momentum after shell formation). At $n_{\rm H}\gtsimm 1\,{\rm cm^{-3}}$, $p_{\rm SF}^{n}$ and $p_{\rm final}$ are both constant when $f_{\rm ML}$ is small, and then decline roughly as a power-law with $f_{\rm ML}$ once $f_{\rm ML}\gtsimm 3$. In contrast, at $n_{\rm H}=0.01\,{\rm cm^{-3}}$, $p_{\rm SF}^{n}$ shows a slight increase with $f_{\rm ML}$ until $f_{\rm ML}\sim 1$. 

Fig.~\ref{fig:varfML_mtm} also shows that when $\nu=10$, the ratio of $p_{\rm final}/p_{\rm SF}^{n}$ increases again when $f_{\rm ML}\gtsimm10$. This is because the average pressure of the ``hot'' gas within the remnant does not drop sharply (by an order of magnitude or so) at the time of shell formation, but rather shows a more gradual reduction. A more gradual reduction occurs also for the case of no mass-loading (see, e.g., Fig.~\ref{fig:nH1fML0_stats}f), and it is this feature which allows $p_{\rm final}/p_{\rm SF}^{n}$ to be significantly greater than unity.

Figs.~\ref{fig:nH0.01varnu_stats}-\ref{fig:nH100varnu_stats} show that $t_{\rm SF}^{n}$ and $r_{\rm SF}^{n}$ display a functional form with $f_{\rm ML}$ that is similar to the form that $p_{\rm SF}^{n}$ and $p_{\rm final}$ have when mass-loading is not limited. A good fit can be obtained to $t_{\rm SF}^{n}$, $r_{\rm SF}^{n}$, $p_{\rm SF}^{n}$ and $p_{\rm final}$ for a specific $n_{\rm H}$ (and high value of $\nu$) using a smoothly broken power-law in $f_{\rm ML}$. The fit function is:
\begin{equation}
\label{eq:brokenPLfit_function1}
A \left\{\frac{1}{2}\left[1 + \left(\frac{f_{\rm ML}}{b}\right)^{1/\Delta}\right]\right\}^{-c\Delta}.
\end{equation}
The results for fitting the $\nu=10^{8}$ data for $t_{\rm SF}^{n}$, $r_{\rm SF}^{n}$, $p_{\rm SF}^{n}$ and $p_{\rm final}$ at specific values of $n_{\rm H}$ are noted in Table~\ref{tab:fit_results_allNH}.

\begin{table}
\centering
\caption[]{Fitted values to $t_{\rm SF}^{n}$, $r_{\rm SF}^{n}$, $p_{\rm SF}^{n}$ and $p_{\rm final}$ using Eq.~\ref{eq:brokenPLfit_function1}. The fit is to the $\nu=10^{8}$ data for specific values of $n_{\rm H}$. Where errors are not given the fit is not well constrained.}
\label{tab:fit_results_allNH}
\begin{tabular}{ccccc}
\hline
      & $t_{\rm SF}^{n}$ & $r_{\rm SF}^{n}$ & $p_{\rm SF}^{n}$ & $p_{\rm final}$ \\
      & (kyr) & (pc) & ($\Msol\,{\rm km\,s^{-1}}$) & ($\Msol\,{\rm km\,s^{-1}}$)\\
\hline
\multicolumn{5}{c}{$n_{\rm H} = 100\,{\rm cm^{-3}}$}\\
$A$ & $2.93\pm0.14$ & $2.78\pm0.04$ & $1.10$ & $1.48\pm0.01$ \\
$b$ & $3.3\pm0.8$ & $3.9\pm0.4$ & $1.96\pm0.24$ & $0.11\pm0.01$ \\
$c$ & $0.40\pm0.04$ & $0.28\pm0.01$ & $0.090\pm0.004$ & $0.092\pm0.001$ \\
$\Delta$ & $0.79\pm0.12$ & $0.94\pm0.05$ & $0.02$ & $0.10\pm0.10$ \\
\hline
\multicolumn{5}{c}{$n_{\rm H} = 1\,{\rm cm^{-3}}$}\\
$A$ & $38.7\pm3.4$ & $19.7\pm0.3$ & $2.25$ & $2.92\pm0.04$ \\
$b$ & $2.97\pm1.22$ & $3.4\pm0.4$ & $1.83\pm0.42$ & $0.38\pm0.06$ \\
$c$ & $0.44\pm0.07$ & $0.28\pm0.01$ & $0.141\pm0.012$ & $0.141\pm0.006$ \\
$\Delta$ & $0.93\pm0.18$ & $1.01\pm0.05$ & $0.057$ & $0.57\pm0.15$ \\
\hline
\multicolumn{5}{c}{$n_{\rm H} = 0.01\,{\rm cm^{-3}}$}\\
$A$ & $581\pm32$ & $141\pm2$ & $4.31$ & $5.52\pm0.14$ \\
$b$ & $2.2\pm0.5$ & $2.8\pm0.3$ & $2.65\pm0.90$ & $3.16\pm0.69$ \\
$c$ & $0.43\pm0.04$ & $0.28\pm0.01$ & $0.159\pm0.022$ & $0.18\pm0.02$ \\
$\Delta$ & $0.61\pm0.15$ & $0.88\pm0.06$ & $0.013$ & $0.20\pm0.20$ \\
\hline
\end{tabular}
\end{table}

\begin{figure*}
\includegraphics[width=8.0cm]{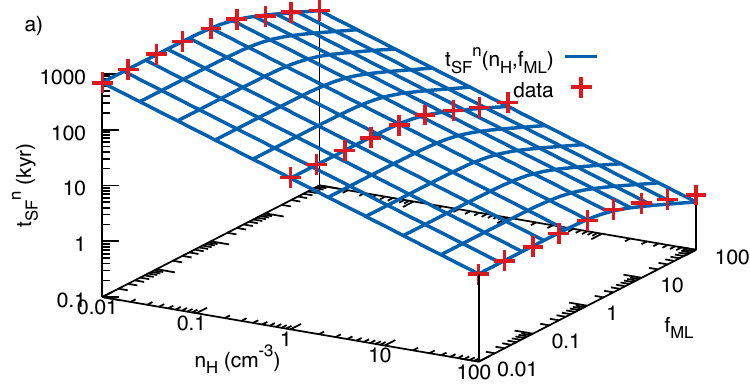}
\includegraphics[width=8.0cm]{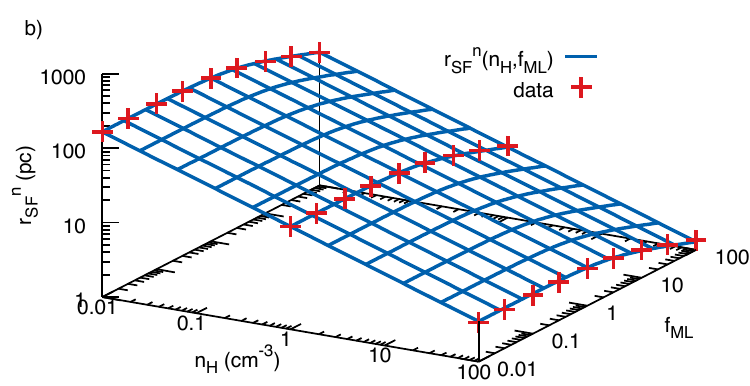}
\includegraphics[width=8.0cm]{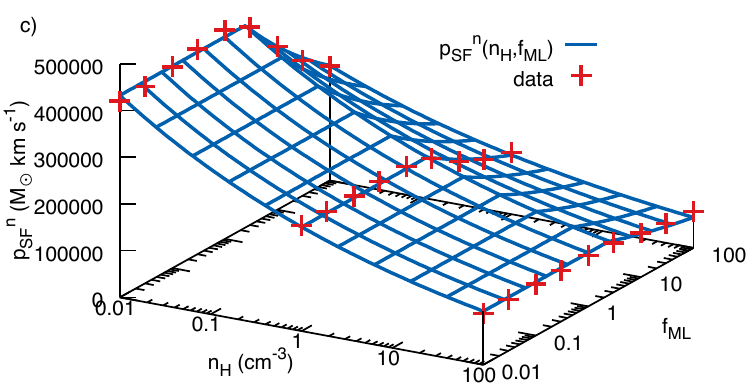}
\includegraphics[width=8.0cm]{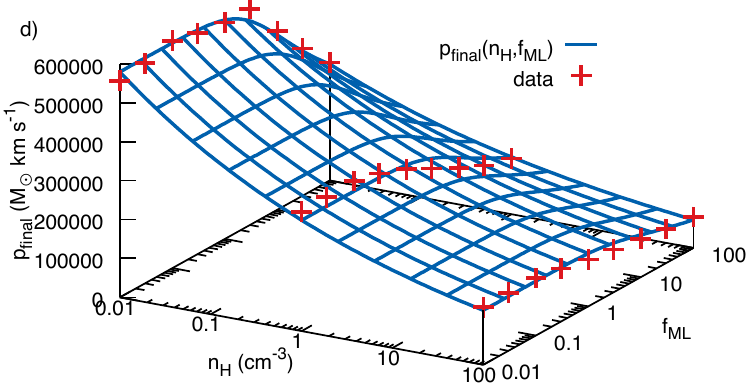}
\caption{Fits to: a) the time of numerical shell formation, $t=t_{\rm SF}^{n}$; b) the remnant radius at $t=t_{\rm SF}^{n}$ ($r_{\rm SF}^{n}$); c) the total radial momentum at $t=t_{\rm SF}^{n}$ ($p_{\rm SF}^{n}$); d) the total radial momentum at $t=3\,t_{\rm SF}^{n}$ ($p_{\rm final}$). The fits are to data from models with $\nu=10^{8}$, and are a function of $f_{\rm ML}$ and $n_{\rm H}$. The surfaces show the fit using Eq.~\ref{eq:brokenPLfit_function} with the parameters in Table~\ref{tab:fit_results}.}
\label{fig:mtm_pSF_nu1e8_fit}
\end{figure*}

A relatively good fit to $t_{\rm SF}^{n}$, $r_{\rm SF}^{n}$, $p_{\rm SF}^{n}$ and $p_{\rm final}$ using a smoothly broken power-law in $f_{\rm ML}$, and a power-law in $n_{\rm H}$, to the case where mass-loading is not limited can also be obtained. The fit function is:
\begin{equation}
\label{eq:brokenPLfit_function}
A \left\{\frac{1}{2}\left[1 + \left(\frac{f_{\rm ML}}{b}\right)^{1/\Delta}\right]\right\}^{-c\Delta} n_{\rm H}^{d}.
\end{equation}
The results for fitting the $\nu=10^{8}$ data for $t_{\rm SF}^{n}$, $r_{\rm SF}^{n}$, $p_{\rm SF}^{n}$ and $p_{\rm final}$ are noted in Table~\ref{tab:fit_results} and plotted in Fig.~\ref{fig:mtm_pSF_nu1e8_fit}. The density exponent for these fits is in good agreement with previous work \citep[cf.][]{Cioffi:1988,Kim:2015,Martizzi:2015}.

\begin{table}
\centering
\caption[]{Fitted values to $t_{\rm SF}^{n}$, $r_{\rm SF}^{n}$, $p_{\rm SF}^{n}$ and $p_{\rm final}$ using Eq.~\ref{eq:brokenPLfit_function}. The fit is to the $\nu=10^{8}$ data.}
\label{tab:fit_results}
\begin{tabular}{ccccc}
\hline
      & $t_{\rm SF}^{n}$ & $r_{\rm SF}^{n}$ & $p_{\rm SF}^{n}$ & $p_{\rm final}$ \\
      & (kyr) & (pc) & ($\Msol\,{\rm km\,s^{-1}}$) & ($\Msol\,{\rm km\,s^{-1}}$)\\
\hline
$A$ & $42.4\pm2.4$ & $20.3\pm0.2$ & $2.22\times10^{5}$ & $2.59\times10^{5}$\\	
$b$ & $2.16\pm0.26$ & $2.79\pm0.16$ & $2.45\pm0.52$ & $2.84\pm1.9$\\ 
$c$ & $0.43\pm0.02$ & $0.28\pm0.006$ & $0.152\pm0.013$ & $0.185\pm0.043$\\
$d$ & $-0.57\pm0.01$ & $-0.42\pm0.001$ & $-0.144\pm0.002$ & $-0.154\pm0.004$ \\
$\Delta$ & $0.61\pm0.07$ & $0.88\pm0.03$ & $0.028\pm32.9$ & $0.74\pm0.41$ \\
\hline
\end{tabular}
\end{table}

\subsection{Multiple SNe}
\label{sec:multSNenoML}

In this subsection we consider the case of multiple SNe that explode into either a homogeneous medium with no mass-loading, or an inhomogeneous environment with mass-loading. We follow the procedure in \citet{Kim:2015} whereby SNe occur at fixed intervals of $0.1$\,Myr. However, rather than considering a range of ambient densities, here we consider only a single ambient density ($n_{\rm H} = 1\,{\rm cm^{-3}}$) and instead vary the mass-loading strength ($f_{\rm ML}$) and the available cloud reservoir ($\nu$). We assume that each SN explosion is co-located. For each event we add $10\,\Msol$ of ejecta and $10^{51}\,$erg of energy ($E_{\rm SN}$). The ejecta is added into a region extending to $0.02\,r_{\rm SF}$. The existing mass, $M_{\rm i}$, and energy, $E_{\rm i}$, within this region is added to the ejecta mass and energy. The total mass is then uniformly distributed. The gas within the ejecta region is assumed to have a thermal temperature of $10^{4}$\,K, and its thermal energy $E_{\rm th}$ is subtracted from the total energy, $E_{\rm tot} = E_{\rm SN} + E_{\rm i}$. A linear velocity profile within the ejecta region is then set so that the kinetic energy of the gas, $E_{\rm kin} = E_{\rm tot} - E_{\rm th}$. This prescription means that the total mass and energy is conserved during this operation, while producing a linear velocity profile for the gas within this region. We run each simulation until 10 SNe have exploded. The last SN occurs at $t=0.9\,$Myr, and the simulation is halted after 2\,Myr. We also calculate a model of a single SN with $E_{\rm SN} = 10^{52}$\,erg for comparison (which we refer to as model nH1fML0E52). We add an ``mSNe'' suffix to model names to indicate those with multiple supernova explosions.

\begin{figure*}
\includegraphics[width=17.5cm]{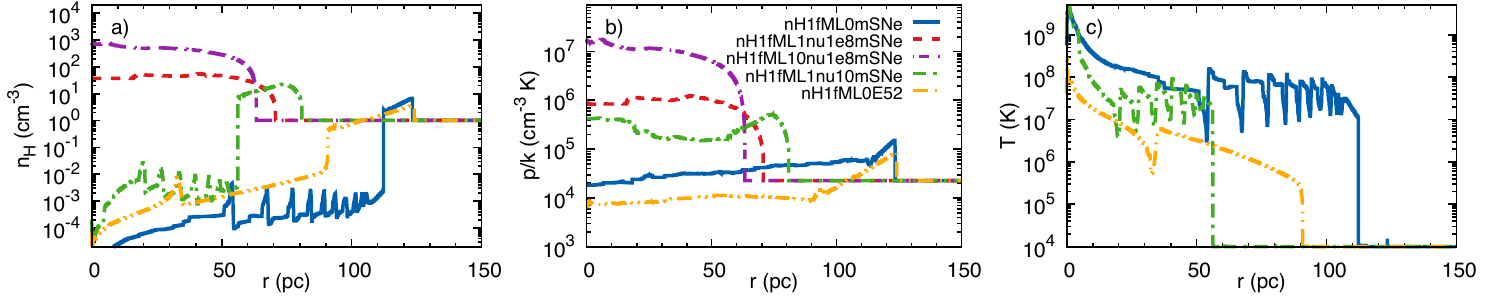}
\caption{Radial profiles at $t=2$\,Myr of models with multiple SNe exploding into an ambient density $n_{\rm H} = 1\,{\rm cm^{-3}}$. The SNe occur at intervals of $0.1$\,Myr and each inject $10^{51}\,$erg of energy and $10\,\Msol$ of mass. Ten SNe explode altogether, with the last explosion occuring at $t=0.9$\,Myr. The models shown are nH1fML0mSNe (no mass-loading), nH1fML1nu1e8mSNe ($f_{\rm ML}=1$, $\nu=10^{8}$), nH1fML10nu1e8mSNe ($f_{\rm ML}=10$, $\nu=10^{8}$), and nH1fML1nu10mSNe ($f_{\rm ML}=1$, $\nu=10$). Also shown for comparison is model nH1fML0E52, a single SN event with $10^{52}\,$erg of energy and $10\,\Msol$ of ejecta with no mass-loading. The panels display: a) number density; b) pressure; c) temperature.}
\label{fig:nH1mSNe_profiles}
\end{figure*}

Fig.~\ref{fig:nH1mSNe_profiles} shows the radial profiles at $t=2$\,Myr of models nH1fML0mSNe (no mass-loading), nH1fML1nu1e8mSNe ($f_{\rm ML}=1$, $\nu=10^{8}$), nH1fML1nu1e8mSNe ($f_{\rm ML}=10$, $\nu=10^{8}$), nH1fML10nu10mSNe ($f_{\rm ML}=1$, $\nu=10$), and nH1fML0E52 (a single SN event of energy $10^{52}$\,erg, with no mass-loading). First we compare models without mass-loading (models nH1fML0mSNe and nH1fML0E52). The forward shock is at roughly the same radius, but this is simply fortuitous timing since the remnant expansion speed is currently significantly higher in the multiple SN model ($v=31\kmps$ versus $v=24\kmps$). The interior structure of the remnant is also quite different. In particular, the multiple SN model has a higher interior pressure and temperature, and a lower interior density. The swept-up shell is also much narrower in the multiple SN model, due to the higher expansion speed that this remnant currently has. Fig.~\ref{fig:nH1mSNe_profiles} also shows profiles for models with mass-loading. In each case the interior density and pressure reflects the mass-loading into the remnants which has come to dominate the evolution by this late time. No hot gas remains in the models with unlimited mass-loading ($\nu=10^{8}$), but some hot gas is present in model nH1fML1nu10mSNe. In this latter model mass-loading has occurred in the region behind the forward shock, but all of the closer in clouds have been destroyed and mass injection has been limited in the interior as a result.

\begin{figure*}
\includegraphics[width=17.5cm]{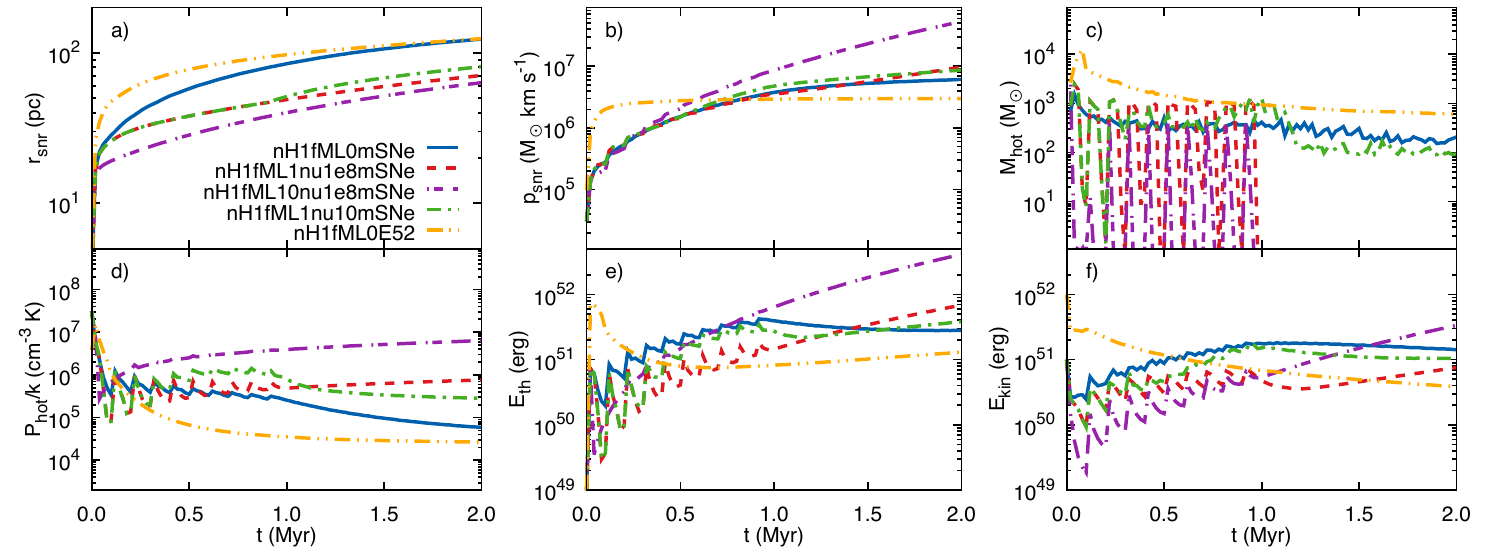}
\caption{The time evolution of models with multiple SNe, with SN intervals of $0.1$\,Myr and $10^{51}\,$erg of energy, into an ambient density $n_{\rm H} = 1\,{\rm cm^{-3}}$. The models shown are nH1fML0mSNe (no mass-loading), nH1fML1nu1e8mSNe ($f_{\rm ML}=1$, $\nu=10^{8}$), nH1fML10nu1e8mSNe ($f_{\rm ML}=10$, $\nu=10^{8}$), and nH1fML1nu10mSNe ($f_{\rm ML}=1$, $\nu=10$). Also shown for comparison is model nH1fML0E52, a single SN event with $10^{52}\,$erg of energy and $10\,\Msol$ of ejecta with no mass-loading. The panels display: a) the radius; b) the total radial momentum; c) the mass of interior ``hot'' gas; d) the pressure of interior ``hot'' gas; e) the thermal energy; f) the kinetic energy.}
\label{fig:nH1variousfMLmSNe_stats}
\end{figure*}

Fig.~\ref{fig:nH1variousfMLmSNe_stats} shows the evolution of various quantities in these models. The discrete injection of kinetic energy and momentum from each SN explosion is clearly visible. Comparing models without mass-loading first, we see that the final momentum in the multiple SN model (model nH1fML0mSNe; $p_{\rm final} = 6.1\times10^{6}\,\Msol\,{\rm km\,s^{-1}}$) exceeds that in the single SN model with the same total energy (model nH1fML0E52; $p_{\rm final} = 3.0\times10^{6}\,\Msol\,{\rm km\,s^{-1}}$) by a factor of two (note also the much more rapid rise in $p_{\rm snr}$ in model nH1fML0E52 - this is due to most of its momentum being created by $t_{\rm SF}^{n} \approx 0.1$\,Myr). The thermal, kinetic and total energy are also all higher in the multiple SN model, though the mass of hot gas is lower. Evidently, in this case multiple SN explosions radiate less energy as subsequent SNe explode into a lower density environment. This allows the remnant to do more PdV work on the swept-up gas, resulting in a higher final momentum. 

However, if mass-loading occurs, subsequent SNe may explode into a denser environment than their predessor(s). This is indeed what we see in models nH1fML1nu1e8mSNe and nH1fML10nu1e8mSNe. The higher density that the SN explosions encounter in models with more rapid mass-loading, and the stronger mass-loading effects that each of them individually experiences, reduces the impact of each explosion, so that the radius of the forward shock grows much more slowly (c.f. models nH1fML0mSNe, nH1fML1nu1e8mSNe and nH1fML10nu1e8mSNe in Fig.~\ref{fig:nH1variousfMLmSNe_stats}a). In fact, in the latter two models the hot gas that is created by each SN explosion largely or completely disappears before the subsequent SN occurs. Thus a reservoir of hot gas is not built up. As we have noted before, so much mass may be injected at the floor temperature of the simulation that the thermal energy of the gas and its associated pressure can drive the final momentum to unrealistically high values (this is the case for model nH1fML10nu1e8mSNe).

On the other hand, if mass-loading runs out because all of the available mass has been used up then subsequent SNe may be able to generate and maintain a hot phase and do more PdV work. This is the case in model nH1fML1nu10mSNe, where explosions from the $3^{\rm rd}$ SN onwards are able to create a long-lasting hot phase in the central region of the remnant (at $t=2$\,Myr this region extends out to a radius of 55\,pc). Divergence with the model with unlimited cloud mass is first seen in the plots of $M_{\rm hot}$, $E_{\rm th}$ and $E_{\rm kin}$ after the $3^{\rm rd}$ explosion, but is not seen in $r_{\rm snr}$ until after the last explosion at $t=0.9$\,Myr.

Table~\ref{tab:mSNe_results} notes some key physical quantities at $t=2$\,Myr for the models discussed in this section. In the absence of mass-loading, the clustered SN model returns more than twice as much energy to the ISM than the single SN model with $E_{\rm SN}=10^{52}$\,erg. 

Our study of just 5 models in this section means that we have only scratched the surface of the rich behaviour that one might expect of mass-loaded clustered SNRs. However, a deeper investigation goes beyond the scope of the present paper and so a more detailed study is left to future work.

\section{Discussion}
\label{sec:discussion}
We first examine the implications of our results for models of momentum-regulated galaxy formation and evolution. We then discuss our chosen parameterization for the mass-loading rate, how our work compares against other work in the literature, and issues associated with sweeping up mass with non-negligible thermal energy and the imposition of a temperature floor. Finally we discuss the limitations of our 1D approach.

\subsection{Implications of low momentum efficiency of mass-loaded SNRs}
The final radial momentum attained per SN is the most important parameter in models of momentum-driven feedback \citep*[e.g.,][]{Hopkins:2014,Kim:2015,Martizzi:2015,Walch:2015}. Non-clustered models of SN feedback typically find that $p_{\rm final} \approx 2-5\times10^{5}\,\Msol\,\kmps$, with a weak dependence on density. In this paper we have shown how mass-loading into SNRs can reduce the final momentum (measured at $t=3\,t_{\rm SF}^{n}$) by about a factor of two (when $f_{\rm ML} \sim 100$ and the cloud mass is not limited). Using the fit results shown in Eq.~\ref{eq:fit_to_pfinal} we find that the final momentum injected into an inhomogeneous medium when $f_{\rm ML} = 100$ is 56 per cent of the amount injected into a uniform medium. If $f_{\rm ML} = 10$ this number is 70 per cent, and if $f_{\rm ML} = 10^{3}$ it is 36 per cent. 

Our results indicate that remnants expanding into an inhomogeneous medium may inject significantly less momentum. In the existing literature there is disagreement on the level of this reduction, with several works finding that there is little difference. This includes \citet{Kim:2015} who find a 5 per cent reduction, \citet{Walch:2015} who find a 7 per cent reduction, and \citet{Iffrig:2015} (who do not specify a particular value, but claim that the final momentum ``does not vary much''). On the other hand, \citet{Martizzi:2015} find a 30 per cent reduction, which they note is due to enhanced radiative losses as shocks pass through the denser clouds in their simulations. Enhanced cooling is also responsible for the momentum reduction that is seen in our simulations, due to the higher densities and lower temperatures that mass-loading creates. 

Ideally we would like to understand the reason(s) for this variation in the reduction. However, there are many differences in the setup and numerics of these works (hereafter referred to as ``the 2015 papers'') which make it difficult to determine the exact cause(s). Differences in the setup include how the ejecta is initialized and the nature and properties of the inhomogeneities/clouds. For example, \citet{Kim:2015} and \citet{Iffrig:2015} inject thermal energy for the SN explosion, while \citet{Walch:2015} inject kinetic energy. \citet{Martizzi:2015} inject 93 per cent of the explosion energy as kinetic energy. It could be argued that the ejecta energy should initially be predominantly kinetic, since the ejecta may well interact with clumps prior to its thermalization.

\citet{Kim:2015} create a 2-phase medium via the thermal instability of an unstable phase with a 10 per cent density perturbation. This creates clouds with density contrasts of $60-90$, and cloud to intercloud mass ratios $\nu\approx5-7$. \citet{Walch:2015} and \citet{Martizzi:2015} set up a lognormal density PDF. \citet{Iffrig:2015} have a more complicated density distribution that has a high-density tail resulting from the gravitational collapse of their cloud. The relatively low reduction found by \citet{Kim:2015} may partially result from their somewhat limited cloud mass. Tables~\ref{tab:ML_results_nu10} and~\ref{tab:ML_results} show that the final momentum is larger for $f_{\rm ML}\gtsimm3$ when $\nu=10$, compared to $\nu=10^{8}$ (an infinite reservoir of cloud mass), indicating a less significant effect once the available cloud mass starts to run out. \citet{Martizzi:2015} commented on the effect of the width of their lognormal density distribution (which was controlled by the value of their turbulent Mach number, $\mathcal{M}$), noting that their results were largely independent of $\mathcal{M}$ when $\mathcal{M}\gtsimm 10$, while a simulation with $\mathcal{M}=1$ produced only a minor reduction in the momentum injection from the uniform medium case. As $\mathcal{M}$ increases, the maximum cloud density contrast increases. However, it is not clear how to relate a log normal (cloud) density distribution to an appropriate value for $\nu$, or how to relate the shock-cloud interactions in these works to an appropriate value for $f_{\rm ML}$ (though see Sec.~\ref{sec:comp_korolev}).

Finally, the 2015 papers use different numerical approaches and codes, and use different numerical resolutions. \citet{Walch:2015} use SPH (SEREN), and initialize the ejecta with 80 particles. The others use grid-based hydrodynamics (RAMESES and Athena). \citet{Martizzi:2015} uses a minimum of 50 grid cells per cooling radius (i.e. the remnant radius at the time of shell formation, $r_{\rm SF}$). \citet{Kim:2015} conduct a detailed convergence study and find that values of $p_{\rm SF}$ are within 5 per cent of each other if there is a minimum of 10 grid cells per shell formation radius, and the initial ejecta radius, $r_{\rm init}$, satisfies $r_{\rm init} < r_{\rm SF}/3$. \citet{Iffrig:2015} use cell widths down to 0.02\,pc in their high resolution calculations. Therefore, \citet{Martizzi:2015} appears to have the best resolved simulations.

One thing that is consistent between the current work and that of \citet{Kim:2015} and \citet{Martizzi:2015} is the finding that the final momentum has a slightly stronger dependence on density in the inhomogeneous case than in the uniform density case (compare Eqs.~\ref{eq:pfinal_fit} and~\ref{eq:fit_to_pfinal}).

To conclude, models that handle subgrid feedback through explicit momentum injection \citep*[e.g.,][]{Kim:2011,Hopkins:2014,Kimm:2015,Goldbaum:2016}, and those that attempt to include SN feedback by explicitly resolving the ST stage \citep[e.g.,][]{Hopkins:2018}, should be re-examined in light of our findings that mass-loading may cause a larger reduction in the final momentum than previously thought.

\subsection{Implications of mass-loading from embedded clouds on clustered SN feedback}
The momentum yield per supernova from clustered supernovae is currently highly uncertain, with some works finding that clustering enhances the yield, and other works finding that it can cause a decrease \citep[see][and references therein]{Gentry:2017}. More recently, \citet{Gentry:2019} investigated the momentum yield for SNe exploding into an ambient density $\rho = 1.33\,m_{\rm H}\,{\rm g\,cm^{-3}}$ (i.e. equivalent to our $n_{\rm H} = 1\,{\rm cm^{-3}}$ simulations). They claim that 1D models provide an upper limit of $3\times10^{6}\,\Msol\,\kmps$ per SN, since such models allow almost no transport of mass and energy across the contact discontinuity, and that 3D models, which overestimate this mixing, provide a lower limit of $2\times10^{5}\,\Msol\,\kmps$ per SN. Thus there remains an order of magnitude uncertainty on the momentum yield per clustered SN. 

Our mass-loading results show that the momentum yield per clustered supernova is affected by mass injection from embedded clouds into the interior of the remnants/superbubble. The long lifetime of superbubbles, and their great range/size, in principle allows significant mass-loading into these structures via this route. However, most superbubble models in the literature consider only mass evaporation from the swept-up shell \citep[e.g.,][]{Sharma:2014,Keller:2014,Gupta:2016,Yadav:2017,Gentry:2019}. Recently, \citet{El-badry:2019} determined that cooling at the shell interface can reduce the shell mass evaporation rate found by \citet{Weaver:1977} by an order of magnitude, so that non-linear mixing now controls the mass transfer into the superbubble interior. The reduction in the mass-loading rate from the shell means that mass-loading from embedded clouds now has a better chance of dominating the overall mass injection into superbubbles. Therefore the effect of mass injection from embedded clouds on superbubble evolution and dynamics deserves more detailed future study.

\subsection{Choice of $f_{\rm ML}$}
The mass-loading rate in our simulations is parameterized by the value of $f_{\rm ML}$. This separates the effect of the mass-loading on the remnant from the uncertainties that currently exist in our understanding of the physics of cloud destruction and mass addition into flows\footnote{For the interaction of a shock with a single cloud see, e.g., \citet{Pittard:2016} and references therein. For the interaction of a wind with a cloud see \citet{Banda-Barragan:2019}. For a direct comparison of wind-cloud and shock-cloud interactions see \citet{Goldsmith:2017} and \citet{Goldsmith:2018}. For the interaction of a flow with many clouds see \citet{Poludnenko:2002} and \citet{Aluzas:2012}.}. It is clear that $f_{\rm ML}$ is determined by the physical situation being studied and will depend on cloud lifetimes, masses, and spatial distributions, and thus will be problem specific (e.g., the cloud spectrum in a starburst galaxy is likely to be different to that in our local ISM).

\subsection{Comparison to \citet{White:1991b}}
\citet{White:1991b} investigated the mass-loading of a remnant by embedded clouds with no radiative losses. In order to obtain a similarity solution they required that the mass injection rate per unit volume varies inversely with time and that its immediate post-shock value is
\begin{equation}
j_{\rm s} = \frac{C}{\tau} \frac{\rho_{0}}{t},
\end{equation} 
where $C$ is the cloud to intercloud mass ratio (equivalent to our $\nu$), and $\tau$ is the ratio of the cloud evaporation time to the remnant age.

Although there are major differences between our approach and that of \citet{White:1991b}, it is still useful to examine their findings and to consider how they relate to ours. Since we do not specify a destruction mechanism for the clouds in our work, we simply consider $\tau$ to be the ratio of the cloud lifetime to the remnant age. We now wish to compare the relationship between \citet{White:1991b}'s $j_{\rm s}$ and our $q$. For $q$ to equal $j_{\rm s}$ at the time of shell formation requires that $\frac{C}{\tau} = \frac{6}{5} f_{\rm ML}$. \citet{White:1991b} note that when $\tau >> C$ the evaporated cloud mass is small and the ST solution is obtained. This is equivalent to $f_{\rm ML} << 1$ in our work. \citet{White:1991b} also find that only when $C \gtsimm 1$ and $1 \ltsimm \tau \ltsimm C$ does the mass-loaded solution differ significantly from the ST solution. This implies that $f_{\rm ML} \gtsimm 5/6$ for $\nu\gtsimm 1$, and is consistent with the values that we need in order to see significant differences (cf. Fig.~\ref{fig:nH1fML0and1_profiles}). Finally, \citet{White:1991b} note that in many cases a simpler 1-parameter solution is adequate where values of $C/\tau$ are varied with $\tau \rightarrow \infty$. The latter requirement is equivalent to us setting $\nu\rightarrow \infty$, and varying the single parameter $f_{\rm ML}$.

\subsection{Comparison to \citet{Korolev:2015}} 
\label{sec:comp_korolev}
\citet{Korolev:2015} performed 2D axisymmetric simulations of a remnant expanding into (toroidal) clouds with a density contrast of 150, and a normal distribution of radii with a mean of 1.5\,pc and a dispersion of 1\,pc. The intercloud density was set to $0.1\,{\rm cm^{-3}}$, and the volume filling factor of the clouds varied between $f=0.05-0.2$. They found that the SNR efficiently destroys close-in clouds via the Kelvin-Helmholtz instability, but that further out and more massive clouds undergo less disruption. Nevertheless, the clouds have an appreciable effect on the global properties of the remnant, slowing its expansion and causing it to cool more quickly. They also find that the onset time of the radiative phase scales roughly as $f^{-1/2}$. We now try to relate this to the scaling that we find in our work when $f_{\rm ML} \gtsimm 2$: $t_{\rm SF} \propto f_{\rm ML}^{-0.43\pm0.02}$ (see Table~\ref{tab:fit_results}).

First, we note that in \citet{Korolev:2015}'s simulations the cloud to intercloud mass ratio, $C$, varies from 7.89 to 37.5. In comparison to our work, we note that our value of $f_{\rm ML}$ cannot exceed $C$ without the cloud mass running out by the time of shell formation. If we assume that the clouds that have been overrun by the onset of the radiative phase are all disrupted, then this implies that $f_{\rm ML} \ltsimm C$. Over the range investigated by \citet{Korolev:2015}, $C\propto f^{1.12}$. If we make the assumption that $f_{\rm ML} \propto C$, we find that $f_{\rm ML} \propto f^{1.12}$. This then implies that $t_{\rm SF} \propto f^{-0.48\pm0.02}$, which is in good agreement with the claim made by \citet{Korolev:2015}.

\subsection{Effect of sweeping up gas with non-negligble thermal energy}
The late-time behaviour of the remnant can be affected if the mass that is swept up contains non-negligible thermal energy. This is the case in our simulations where we adopt $T_{\rm amb}=10^{4}$\,K. Reducing the temperature of the ambient medium to $T_{\rm amb}=10^{2}$\,K causes the final momentum obtained in models without mass-loading to be 12 per cent lower at $t=10\,t_{\rm SF}$ (see Appendix~\ref{sec:tfloor}).

\subsection{Effect of a temperature floor}
In some of our heavily mass-loaded models the late time behaviour of the radial momentum is affected by our imposition of a floor temperature at $10^{4}$\,K. This can result in some of the mass being injected with non-zero thermal energy. Though this behaviour is numerical in origin, it is not necessarily unphysical. Dense atomic or molecular gas in the clouds may well be heated to $10^{4}$\,K, perhaps by ionizing photons, as it mixes into the environment. Appendix~\ref{sec:tfloor} shows that setting a temperature floor of $10^{2}$\,K removes most of this source of extra energy in our simulations.

\subsection{Limitations of 1D models}
\label{sec:limitations}
Our 1D models are of course limited by the assumptions of spherical symmetry and a continuous approximation for the clouds. In reality each cloud is a discrete object and will locally perturb the flow, and may individually or collectively cause global asymmetries, such as the reverse shock no longer converging at the centre of the remnant and the formation time and radius of the shell differing in different parts of the remnant. Remnants expanding into an inhomogeneous medium will therefore always have some level of asymmetry. Such asymmetry will be at a maximum when the remnant encounters very few, but very big and dense clouds, and will be reduced to a minimum when many small clouds are encountered. If one is interested in the details of such asymmetries, or in specific local conditions, then 3D calculations become necessary.

\section{Summary and conclusions}
\label{sec:summary}
We present the results of 1D spherically symmetric simulations of SNRs interacting with a clumpy, inhomogeneous medium. In all cases the SN explosion deposits $10\,\Msol$ of mass and $10^{51}$\,erg of kinetic energy into the environment, within a volume of radius $0.02\,r_{\rm SF}$. The expanding remnant is assumed to sweep over pre-existing clouds which are destroyed within it as they become overrun/engulfed. This destruction adds mass into the remnant and affects its behaviour and evolution. The nature of the mass-loading is parameterized using two variables: its strength depends on the value of $f_{\rm ML}$ (values $\gtsimm 1$ indicate that mass-loading is significant on the timescale of shell formation), while its duration depends on the value of $\nu$ (the ratio of cloud to inter-cloud mass per unit volume in the ambient medium). The mass injection within the remnant is assumed to occur at a uniform rate per unit volume, unless and until the available mass reservoir at a particular radius is exhausted.

We find that:
\begin{enumerate}
\item Mass-loading into the remnant can affect the behaviour and evolution of the remnant from its earliest stages. Mass-loading drag-heats the ejecta prior to its thermalization at the reverse shock, and in extreme cases can prevent the reverse shock from moving all the way back to the point of explosion.
\item More generally, mass-loading increases the density and pressure, and decreases the temperature and velocity, within the remnant. The remnant does not expand as rapidly or as far. Prior to shell formation, a mass-loaded remnant may contain more hot gas and have a higher radial momentum. However, the remnant also cools more quickly, does less PdV work on the swept-up gas, and ultimately attains a lower final momentum. 
\item Some of the properties of mass-loaded remnants that have an unlimited supply of cloud mass that can be injected into them can be fitted with a broken power-law in $f_{\rm ML}$. We obtain the following fits as a function of $n_{\rm H}$ and $f_{\rm ML}$:
\begin{equation}
\label{eq:fit_to_tSF}
t_{\rm SF}^{n} = 42.4 \left\{\frac{1}{2}\left[1 + \left(\frac{f_{\rm ML}}{2.16}\right)^{1.64}\right]\right\}^{-0.26} n_{\rm H}^{-0.57}\,\,{\rm kyr}.
\end{equation}
\begin{equation}
\label{eq:fit_to_rSF}
r_{\rm SF}^{n} = 20.3 \left\{\frac{1}{2}\left[1 + \left(\frac{f_{\rm ML}}{2.79}\right)^{1.14}\right]\right\}^{-0.25} n_{\rm H}^{-0.42}\,\,{\rm pc}.
\end{equation}
\begin{equation}
\label{eq:fit_to_pSF}
p_{\rm SF}^{n} = 2.22\times10^{5} \left\{\frac{1}{2}\left[1 + \left(\frac{f_{\rm ML}}{2.45}\right)^{35.7}\right]\right\}^{-0.0043} n_{\rm H}^{-0.14}\,\,\Msol\,\kmps.
\end{equation}
\begin{equation}
\label{eq:fit_to_pfinal}
p_{\rm final} = 2.59\times10^{5} \left\{\frac{1}{2}\left[1 + \left(\frac{f_{\rm ML}}{2.84}\right)^{1.35}\right]\right\}^{-0.14} n_{\rm H}^{-0.15}\,\,\Msol\,\kmps.
\end{equation}
\item For large values of $f_{\rm ML}$ and $\nu$, we find that $t_{\rm SF}^{n} \propto f_{\rm ML}^{-0.43\pm0.02}$, $r_{\rm SF}^{n} \propto f_{\rm ML}^{-0.28\pm0.01}$, $p_{\rm SF}^{n} \propto f_{\rm ML}^{-0.15\pm0.01}$, and $p_{\rm final} \propto f_{\rm ML}^{-0.19\pm0.04}$.
\item The final momentum injected into an inhomogeneous medium when $\nu$ is large and $f_{\rm ML} = 100$ is 56 per cent of the amount injected into a uniform medium. If $f_{\rm ML} = 10$ this number is 0.70, and if $f_{\rm ML} = 10^{3}$ it is 0.36. Our results indicate that remnants expanding into an inhomogeneous medium may inject significantly less momentum than previous results in the literature have found. 
\item The evolution of remnants that experience {\em very} rapid (e.g., $f_{\rm ML} \gtsimm 100$) but limited (e.g., $\nu \ltsimm 100$) mass-loading is akin to that of remnants expanding into a uniform medium of the same total smeared-out density. However, we do not genereally expect such extreme rates of mass-loading to occur. When mass-loading occurs at a more measured pace throughout the remnant, the remnant evolves differently to a remnant evolving in the same average ambient density. Thus, in general, the mass-loaded remnants do not follow the scaling with mean density found by \citet{Kim:2015} if the gas injected by clouds effectively mixes.
\item Multiple SN explosions into a homogeneous environment produce higher final momentum per explosion than single explosions of the same total energy if hot gas is maintained between explosions. However, when there are multiple SN explosions into an inhomogeneous environment, subsequent explosions may encounter higher densities than prior explosions due to the liberation of mass from engulfed clouds. If the rate of mass-loading is high and sustained, the hot gas created by each explosion may completely cool prior to the occurance of the next explosion. This reduces the final momentum per explosion compared to the case without mass-loading. In cases where the available amount of cloud mass is finite, later SNe may be able to create a sustained hot phase when earlier SNe have not been able to.
\end{enumerate}

In summary, mass-loading can significantly affect the behaviour of SNe and clustered SNe. The final momentum that each SN delivers to the ISM is a complicated function of the rate at which mass-loading occurs, the amount of available cloud mass, and the effect of previous SNe (including on the reservoir of cloud mass).

\section*{Acknowledgements}
We thank Katie Baker for running some simulations in the very early stages of this work and the referee for some useful suggestions. We acknowledge support from the Science and Technology Facilities Council (STFC, Research Grant ST/P00041X/1). The calculations herein were performed on the DiRAC 1 Facility at Leeds jointly funded by STFC, the Large Facilities Capital Fund of BIS and the University of Leeds and on other facilities at the University of Leeds. Data for the figures in this paper are available from \url{https://doi.org/XXX}.





\begin{thebibliography}{99}
\bibitem[\protect\citeauthoryear{Agertz et al.}{2013}]{Agertz:2013}
Agertz O., Kravtsov A.V., Leitner S.N., Gnedin N.Y., 2013, ApJ, 770, 25
\bibitem[\protect\citeauthoryear{Aguirre et al.}{2001}]{Aguirre:2001}
Aguirre A., Hernquist L., Schaye J., Katz N., Weinberg D.H., Gardner J., 2001, ApJ, 561, 521
\bibitem[\protect\citeauthoryear{Aluzas et al.}{2012}]{Aluzas:2012}
Al$\bar{\rm u}$zas R., Pittard J.M., Hartquist T.W., Falle S.A.E.G., Langton R., 2012, MNRAS, 425, 2212
\bibitem[\protect\citeauthoryear{Anders \& Grevesse}{1989}]{Anders:1989}
Anders E., Grevesse N., 1989, GeCoA, 53, 197
\bibitem[\protect\citeauthoryear{Arthur \& Henney}{1996}]{Arthur:1996}
Arthur S.J., Henny W.J., 1996, ApJ, 457, 752
\bibitem[\protect\citeauthoryear{Banda-Barragan et al.}{2019}]{Banda-Barragan:2019}
Banda-Barrag\'{a}n W.E., Zertuche F.J., Federrath C., Del Valle J.G., Br\"{u}ggen M., Wagner A.Y., 2019, MNRAS, in press
\bibitem[\protect\citeauthoryear{Behroozi, Conroy \& Wechsler}{Behroozi et al.}{2010}]{Behroozi:2010}
Behroozi P.S., Conroy C., Wechsler R.H., 2010, ApJ, 717, 379
\bibitem[\protect\citeauthoryear{Blondin et al.}{1998}]{Blondin:1998}
Blondin J.M., Wright E.B., Borkowski K.J., Reynolds S.P., 1998, ApJ, 500, 342 
\bibitem[\protect\citeauthoryear{Brook et al.}{2012}]{Brook:2012}
Brook C.B., Stinson G., Gibson B.K., Wadsley J., Quinn T., 2012, MNRAS, 424, 1275
\bibitem[\protect\citeauthoryear{Chen, Liu \& Wang}{Chen et al.}{1995}]{Chen:1995}
Chen Y., Liu N., Wang Z.-R., 1995, ApJ, 446, 755
\bibitem[\protect\citeauthoryear{Chevalier \& Clegg}{1985}]{Chevalier:1985}
Chevalier R.A., Clegg A.W., 1985, Nature, 317, 44 
\bibitem[\protect\citeauthoryear{Chieze \& Lazareff}{1981}]{Chieze:1981}
Chieze J.P., Lazareff B., 1981, A\&A, 95, 194
\bibitem[\protect\citeauthoryear{Cioffi, McKee \& Bertschinger}{Cioffi et al.}{1988}]{Cioffi:1988}
Cioffi D.F., McKee C.F., Bertschinger E., 1988, ApJ, 334, 252
\bibitem[\protect\citeauthoryear{Cowie, McKee \& Ostriker}{Cowie et al.}{1981}]{Cowie:1981}
Cowie L.L., McKee C.F., Ostriker J.P., 1981, 247, 908
\bibitem[\protect\citeauthoryear{de Avillez \& Breitschwerdt}{2004}]{deAvillez:2004}
de Avillez M., Breitschwerdt D., 2004, Ap\&SS, 289, 479
\bibitem[\protect\citeauthoryear{Dib, Bell \& Burkert}{Dib et al.}{2006}]{Dib:2006}
Dib S., Bell E., Burkert A., 2006, ApJ, 638, 797
\bibitem[\protect\citeauthoryear{Dyson, Arthur \& Hartquist}{Dyson et al.}{2002}]{Dyson:2002}
Dyson J.E., Arthur S.J., Hartquist T.W., 2002, A\&A, 390, 1063
\bibitem[\protect\citeauthoryear{Dyson \& Hartquist}{1987}]{Dyson:1987}
Dyson J.E., Hartquist T.W., 1987, MNRAS, 228, 453
\bibitem[\protect\citeauthoryear{El-badry et al.}{2019}]{El-badry:2019}
El-Badry K., Ostriker E.C., Kim C.-G., Quataert E., Weisz D.R., 2019, arXiv:1902.09547
\bibitem[\protect\citeauthoryear{Falle}{1975}]{Falle:1975}
Falle S.A.E.G., 1975, MNRAS, 172, 55
\bibitem[\protect\citeauthoryear{Faucher-Giguere, Keres \& Ma}{Faucher-Giguere et al.}{2011}]{Faucher-Giguere:2011}
Faucher-Gigu\`{e}re C.-A., Keres D., Ma C.-P., 2011, MNRAS, 417, 2982
\bibitem[\protect\citeauthoryear{Gatto et al.}{2015}]{Gatto:2015}
Gatto A., et al., 2015, MNRAS, 449, 1057
\bibitem[\protect\citeauthoryear{Gentry et al.}{2017}]{Gentry:2017}
Gentry E.S., Krumholz M.R., Dekel A., Madau P., 2017, MNRAS, 465, 2471
\bibitem[\protect\citeauthoryear{Gentry et al.}{2019}]{Gentry:2019}
Gentry E.S., Krumholz M.R., Madau P., Lupi A., 2019, MNRAS, 483, 3647
\bibitem[\protect\citeauthoryear{Goldbaum, Krumholz \& Forbes}{Goldbaum et al.}{2016}]{Goldbaum:2016}
Goldbaum N.J., Krumholz M.R., Forbes J.C., 2016, ApJ, 827, 28
\bibitem[\protect\citeauthoryear{Goldsmith \& Pittard}{2017}]{Goldsmith:2017}
Goldsmith K.J.A, Pittard J.M., 2017, MNRAS, 470, 2427
\bibitem[\protect\citeauthoryear{Goldsmith \& Pittard}{2018}]{Goldsmith:2018}
Goldsmith K.J.A, Pittard J.M., 2018, MNRAS, 476, 2209
\bibitem[\protect\citeauthoryear{Governato et al.}{2007}]{Governato:2007}
Governato F., Willman B., Mayer L., Brooks A., Stinson G., Valenzuela O., Wadsley J., Quinn T., 2007, MNRAS, 374, 1479 
\bibitem[\protect\citeauthoryear{Grevesse et al.}{2010}]{Grevesse:2010}
Grevesse N., Asplund M., Sauval A., Scott P., 2010, Ap\&SS, 328, 179
\bibitem[\protect\citeauthoryear{Guedes et al.}{2011}]{Guedes:2011}
Guedes J., Callegari S., Madau P., Mayer L., 2011, ApJ, 742, 76
\bibitem[\protect\citeauthoryear{Gupta et al.}{2018}]{Gupta:2018}
Gupta S., Nath B.B., Sharma P., Eichler D., 2018, MNRAS, 473, 1537
\bibitem[\protect\citeauthoryear{Gupta et al.}{2016}]{Gupta:2016}
Gupta S., Nath B.B., Sharma P., Shchekinov Y., 2016, MNRAS, 462, 4532
\bibitem[\protect\citeauthoryear{Hartquist et al.}{1986}]{Hartquist:1986}
Hartquist T.W., Dyson J.E., Pettini M., Smith L.J., 1986, MNRAS, 221, 715
\bibitem[\protect\citeauthoryear{Hennebelle \& Iffrig}{2014}]{Hennebelle:2014}
Hennebelle P., Iffrig O., 2014, A\&A, 570, A81
\bibitem[\protect\citeauthoryear{Hill et al.}{2012}]{Hill:2012}
Hill A.S., Joung M.R., Mac Low M.-M., et al., 2012, ApJ, 750, 104
\bibitem[\protect\citeauthoryear{Hopkins et al.}{2014}]{Hopkins:2014}
Hopkins P.F., Keres D., Onorbe J., Faucher-Giguere C.-A., Quataert E., Murray N., Bullock J.S., 2014, MNRAS, 445, 581
\bibitem[\protect\citeauthoryear{Hopkins, Quataert \& Murray}{Hopkins et al.}{2011}]{Hopkins:2011}
Hopkins P.F., Quataert E., Murray N., 2011, MNRAS, 417, 950
\bibitem[\protect\citeauthoryear{Hopkins, Quataert \& Murray}{Hopkins et al.}{2012}]{Hopkins:2012}
Hopkins P.F., Quataert E., Murray N., 2012, MNRAS, 421, 3522
\bibitem[\protect\citeauthoryear{Hopkins et al.}{2018}]{Hopkins:2018}
Hopkins P.F., et al., 2018, MNRAS, 477, 1578
\bibitem[\protect\citeauthoryear{Iffrig \& Hennebelle}{2015}]{Iffrig:2015}
Iffrig O., Hennebelle P., 2015, A\&A, 576, A95
\bibitem[\protect\citeauthoryear{Katz}{1992}]{Katz:1992}
Katz N., 1992, ApJ, 391, 502
\bibitem[\protect\citeauthoryear{Keller et al.}{2014}]{Keller:2014}
Keller B.W., Wadsley J., Benincasa S.M., Couchman H.M.P., 2014, MNRAS, 442, 3013
\bibitem[\protect\citeauthoryear{Keres et al.}{2009}]{Keres:2009}
Keres D., Katz N., Dav\'{e} R., Fardal M., Weinberg D.H., 2009, MNRAS, 396, 2332
\bibitem[\protect\citeauthoryear{Kim, Kim \& Ostriker}{Kim et al.}{2011}]{Kim:2011}
Kim C.-G., Kim W.-T., Ostriker E.C., 2011, ApJ, 743, 25
\bibitem[\protect\citeauthoryear{Kim \& Ostriker}{2015}]{Kim:2015}
Kim C.-G., Ostriker E.C., 2015, ApJ, 802, 99
\bibitem[\protect\citeauthoryear{Kimm \& Cen}{2014}]{Kimm:2014}
Kimm T., Cen R., 2014, ApJ, 788, 121 
\bibitem[\protect\citeauthoryear{Kimm et al.}{2015}]{Kimm:2015}
Kimm T., Cen R., Devriendt J., Dubois Y., Slyz A., 2015, MNRAS, 451, 2900
\bibitem[\protect\citeauthoryear{Korolev et al.}{2015}]{Korolev:2015}
Korolev V.V., Vasiliev E.O., Kovalenko I.G., Shchekinov Y.A., 2015, ARep, 59, 690
\bibitem[\protect\citeauthoryear{Mac Low \& Klessen}{2004}]{MacLow:2004}
Mac Low M.-M., Klessen R.S., 2004, Reviews of Modern Physics, 76, 125
\bibitem[\protect\citeauthoryear{Martizzi, Faucher-Gigu\`{e}re \& Quataert}{Martizzi et al.}{2015}]{Martizzi:2015}
Martizzi D., Faucher-Gigu\`{e}re C.-A., Quataert E., 2015, MNRAS, 450, 504
\bibitem[\protect\citeauthoryear{McKee \& Ostriker}{1977}]{McKee:1977}
McKee C.F., Ostriker J.P., 1977, ApJ, 218, 148
\bibitem[\protect\citeauthoryear{Moster et al.}{2010}]{Moster:2010}
Moster B.P., Somerville R.S., Maulbetsch C., van den Bosch F.C., Maccio A.V., Naab T., Oser L., 2010, ApJ, 710, 903
\bibitem[\protect\citeauthoryear{Oppenheimer \& Dav\'{e}}{2006}]{Oppenheimer:2006}
Oppenheimer B.D., Dav\'{e} R., 2006, MNRAS, 373, 1265
\bibitem[\protect\citeauthoryear{Ostriker \& Shetty}{2011}]{Ostriker:2011}
Ostriker E.C., Shetty R., 2011, ApJ, 731, 41
\bibitem[\protect\citeauthoryear{Padoan et al.}{2016}]{Padoan:2016}
Padoan P., Pan L., Haugbolle T., Nordlund A., 2016, ApJ, 822, 11
\bibitem[\protect\citeauthoryear{Pittard}{2007}]{Pittard:2007}
 Pittard J.M., 2007, in ``Diffuse Matter from Star Forming Regions to Active Galaxies - A Volume Honouring John Dyson'', eds. T.W. Hartquist, J.M. Pittard and S.A.E.G. Falle, Astrophysics and Space Science Proceedings, Springer Dordrecht, p.245
\bibitem[\protect\citeauthoryear{Pittard et al.}{2003}]{Pittard:2003}
Pittard J.M., Arthur S.J., Dyson J.E., Falle S.A.E.G., Hartquist T.W., Knight M.I., Pexton M., 2003, A\&A, 401, 1027
\bibitem[\protect\citeauthoryear{Pittard, Dyson \& Hartquist}{Pittard et al.}{2001}]{Pittard:2001}
Pittard J.M., Dyson J.E., Hartquist T.W., 2001, Ap\&SS, 278, 269
\bibitem[\protect\citeauthoryear{Pittard, Hartquist \& Falle}{Pittard et al.}{2010}]{Pittard:2010}
Pittard J.M., Hartquist T.W., Falle S.A.E.G., 2010, MNRAS, 405, 821
\bibitem[\protect\citeauthoryear{Pittard \& Parkin}{2016}]{Pittard:2016}
Pittard J.M., Parkin E.R., 2016, MNRAS, 457, 4470
\bibitem[\protect\citeauthoryear{Poludnenko, Frank \& Blackman}{Poludnenko et al.}{2002}]{Poludnenko:2002}
Poludnenko A.Y., Frank A., Blackman E.G., 2002, ApJ, 576, 832
\bibitem[\protect\citeauthoryear{Scannapieco et al.}{2006}]{Scannapieco:2006}
Scannapieco C., Tissera P.B., White S.D.M., Springel V., 2006, MNRAS, 371, 1125
\bibitem[\protect\citeauthoryear{Scannapieco et al.}{2012}]{Scannapieco:2012}
Scannapieco C., et al., 2012, MNRAS, 432, 1726
\bibitem[\protect\citeauthoryear{Sharma et al.}{2014}]{Sharma:2014}
Sharma P., Roy A., Nath B., Shchekinov Y., 2014, MNRAS, 443, 3463
\bibitem[\protect\citeauthoryear{Shen, Wadsley \& Stinson}{Shen et al.}{2010}]{Shen:2010}
Shen S., Wadsley J., Stinson G., 2010, MNRAS, 407, 1581
\bibitem[\protect\citeauthoryear{Shetty \& Ostriker}{2012}]{Shetty:2012}
Shetty R., Ostriker E.C., 2012, ApJ, 754, 2
\bibitem[\protect\citeauthoryear{Slavin et al.}{2017}]{Slavin:2017}
Slavin J.D., Smith R.K., Foster A., Winter H.D., Raymond J.C., Slane P.O., Yamaguchi H., 2017, ApJ, 846, 77
\bibitem[\protect\citeauthoryear{Springel \& Hernquist}{2003}]{Springel:2003}
Springel V., Hernquist L., 2003, MNRAS, 339, 289
\bibitem[\protect\citeauthoryear{Stinson et al.}{2006}]{Stinson:2006}
Stinson G., et al., 2006, MNRAS, 373, 1074
\bibitem[\protect\citeauthoryear{Stinson et al.}{2013}]{Stinson:2013}
Stinson G., et al., 2013, MNRAS, 428, 129
\bibitem[\protect\citeauthoryear{Strickland \& Heckman}{2009}]{Strickland:2009}
Strickland D.K., Heckman T.M., 2009, ApJ, 697, 2030
\bibitem[\protect\citeauthoryear{Sutherland \& Dopita}{1993}]{Sutherland:1993}
Sutherland R.S, Dopita M.A., 1993, ApJSS, 88, 253
\bibitem[\protect\citeauthoryear{Teyssier et al.}{2013}]{Teyssier:2013}
Teyssier R., Pontzen A., Dubois Y., Read J.I., 2013, MNRAS, 429, 3068
\bibitem[\protect\citeauthoryear{Thacker \& Couchman}{2000}]{Thacker:2000}
Thacker R.J., Couchman H.M.P., 2000, ApJ, 545, 728
\bibitem[\protect\citeauthoryear{Thornton et al.}{1998}]{Thornton:1998}
Thornton K., Gaudlitz M., Janka H.-Th., Steinmetz M., 1998, ApJ, 500, 95
\bibitem[\protect\citeauthoryear{Townsend}{2009}]{Townsend:2009}
Townsend R.H.D., 2009, ApJSS, 181, 391
\bibitem[\protect\citeauthoryear{Veilleux, Cecil \& Bland-Hawthorn}{Veilleux et al.}{2005}]{Veilleux:2005}
Veilleux S., Cecil G., Bland-Hawthorn J., 2005, ARA\&A, 43, 769
\bibitem[\protect\citeauthoryear{Vogelsberger et al.}{2013}]{Vogelsberger:2013}
Vogelsberger M., Genel S., Sijacki D., Torrey P., Springel V., Hernquist L., 2013, MNRAS, 436, 3031
\bibitem[\protect\citeauthoryear{Walch \& Naab}{2015}]{Walch:2015}
Walch S., Naab T., 2015, MNRAS, 451, 2757
\bibitem[\protect\citeauthoryear{Wang et al.}{2014}]{Wang:2014}
Wang Y., Ferland G.J., Lykins M.L., Porter R.L., van Hoof P.A.M., Williams R.J.R., 2014, MNRAS, 440, 3100
\bibitem[\protect\citeauthoryear{Weaver et al.}{1977}]{Weaver:1977}
Weaver R., McCray R., Castor J., Shapiro P., Moore R., 1977, ApJ, 218, 377
\bibitem[\protect\citeauthoryear{White \& Frenk}{1991}]{White:1991}
White S.D.M., Frenk C.S., 1991, ApJ, 379, 52
\bibitem[\protect\citeauthoryear{White \& Long}{1991}]{White:1991b}
White R.L., Long K.S., 1991, ApJ, 373, 543
\bibitem[\protect\citeauthoryear{Wiersma et al.}{2010}]{Wiersma:2010}
Wiersma R.P.C., Schaye J., Dalla Vecchia C., Booth C.M., Theuns T., Aguirre A., 2010, MNRAS, 409, 132
\bibitem[\protect\citeauthoryear{Wolff \& Durisen}{1987}]{Wolff:1987}
Wolff M.T., Durisen R.H., 1987, MNRAS, 224, 701
\bibitem[\protect\citeauthoryear{Yadav et al.}{2017}]{Yadav:2017}
Yadav N., Mukherjee D., Sharma P., Nath B.B., 2017, MNRAS, 465, 1720
\bibitem[\protect\citeauthoryear{Zhang \& Chevalier}{2019}]{Zhang:2019a}
Zhang G.-Y., Chevalier R.A., 2019, MNRAS, 482, 1602
\bibitem[\protect\citeauthoryear{Zhang et al.}{2019}]{Zhang:2019b}
Zhang G.-Y., Slavin J.D., Foster A., Smith R.K., ZuHone J.A., Zhou P., Chen Y., 2019, arXiv:1902.10718
\end{thebibliography}




\appendix

\section{Data tables and figures}
\label{sec:appA}
In this appendix we provide data tables and figures for most of the models that are presented in this work. In the data tables we note the values of key physical quantities at the time of numerical shell formation ($t_{\rm SF}^{n}$). We also note values at $t=2$\,Myr for the multiple SN models and the high energy single SN comparison model. The figures show how the key physical quantities in the single mass-loaded SN models vary with $f_{\rm ML}$ and $\nu$ for a fixed intercloud ambient density $n_{\rm H}$.

\begin{table*}
\centering
\caption[]{Physical quantities at the time of numerical shell formation ($t_{\rm SF}^{n}$; columns 4-9) and at $t=3\,t_{\rm SF}^{n}$ (columns 10-12) in models of SNRs expanding into an inhomogeneous ambient medium of intercloud density $n_{\rm H}$. The initial mass of gas in clouds is equal to the intercloud mass ($\nu=1$). The injection of cloud gas into the remnant occurs at a rate characterized by the value of $f_{\rm ML}$.}
\label{tab:ML_results_nu1}
\begin{tabular}{cccccccccccc}
\hline
Model & $n_{\rm H}$ & $f_{\rm ML}$ & $t_{\rm SF}^{n}$ & $r_{\rm SF}^{n}$ & $M_{\rm sw}$ & $M_{\rm inj}$ & $M_{\rm hot}$ & $p_{\rm SF}^{n}/10^{5}$ & $p_{\rm final}/10^{5}$ & $E_{\rm th}/10^{50}$ & $E_{\rm kin}/10^{50}$ \\
      & (cm$^{-3}$) &            & (kyr)           & (pc)           & ($\Msol$)   & ($\Msol$)    & ($\Msol$)   & ($\Msol\,{\rm km\,s^{-1}}$) & ($\Msol\,{\rm km\,s^{-1}}$) & (erg) & (erg) \\
\hline
nH0.01fML0.01nu1  & 0.01 & 0.01 & 690  & 166  & 6400  & 79   & 6480  & 4.20 & 5.56 & 1.8 & 1.7 \\
nH0.01fML0.03nu1  & 0.01 & 0.03 & 690  & 166  & 6400  & 240  & 6630  & 4.22 & 5.66 & 1.8 & 1.7 \\
nH0.01fML0.1nu1   & 0.01 & 0.1  & 710  & 166  & 6400  & 825  & 7260  & 4.31 & 5.85 & 1.5 & 1.5 \\
nH0.01fML0.3nu1   & 0.01 & 0.3  & 690  & 163  & 6100  & 2310 & 8420  & 4.39 & 5.74 & 1.3 & 1.3 \\
nH0.01fML1nu1     & 0.01 & 1    & 640  & 155  & 5240  & 4280 & 9530  & 4.46 & 5.81 & 1.4 & 1.7 \\
nH0.01fML3nu1     & 0.01 & 3    & 550  & 140  & 3830  & 3620 & 7450  & 4.22 & 5.61 & 1.8 & 1.8 \\
nH0.01fML10nu1    & 0.01 & 10   & 400  & 120  & 2390  & 2310 & 4710  & 3.56 & 5.10 & 2.0 & 2.0 \\
nH0.01fML30nu1    & 0.01 & 30   & 460  & 123  & 2630  & 2610 & 5250  & 3.79 & 5.09 & 1.6 & 1.8 \\
nH0.01fML100nu1   & 0.01 & 100  & 445  & 122  & 2550  & 2550 & 5110  & 3.75 & 5.08 & 1.7 & 1.9 \\
nH1fML0.01nu1     & 1    & 0.01 & 51.8 & 24.1 & 1950  & 20   & 1980  & 2.26 & 3.06 & 1.3 & 1.9 \\
nH1fML0.03nu1     & 1    & 0.03 & 51.8 & 24.0 & 1940  & 60   & 2010  & 2.26 & 3.08 & 1.1 & 1.8 \\
nH1fML0.1nu1      & 1    & 0.1  & 50.5 & 23.8 & 1890  & 190  & 2080  & 2.26 & 3.10 & 0.70& 1.7 \\
nH1fML0.3nu1      & 1    & 0.3  & 49.2 & 23.4 & 1790  & 525  & 2330  & 2.28 & 2.99 & 0.33& 1.3 \\
nH1fML1nu1        & 1    & 1    & 45.5 & 22.3 & 1550  & 1140 & 2700  & 2.27 & 2.98 & 0.58& 2.0 \\
nH1fML3nu1        & 1    & 3    & 39.9 & 20.2 & 1160  & 1070 & 2240  & 2.19 & 2.98 & 1.2 & 1.9 \\
nH1fML10nu1       & 1    & 10   & 31.9 & 17.6 & 760   & 730  & 1500  & 1.97 & 2.78 & 1.6 & 2.1 \\
nH1fML30nu1       & 1    & 30   & 35.7 & 18.0 & 820   & 815  & 1650  & 2.07 & 2.75 & 1.1 & 1.9 \\
nH1fML100nu1      & 1    & 100  & 34.7 & 17.9 & 800   & 800  & 1610  & 2.03 & 2.75 & 1.3 & 1.9 \\
nH1fML1000nu1     & 1    & 1000 & 34.7 & 17.9 & 800   & 800  & 1610  & 2.03 & 2.75 & 1.3 & 1.9 \\
nH100fML0.01nu1   & 100  & 0.01 & 3.68 & 3.33 & 515   & 4.1  & 525   & 1.10 & 1.47 & 0.82& 1.8 \\
nH100fML0.03nu1   & 100  & 0.03 & 3.67 & 3.33 & 515   & 12   & 535   & 1.10 & 1.49 & 0.73& 1.7 \\
nH100fML0.1nu1    & 100  & 0.1  & 3.58 & 3.29 & 500   & 39   & 540   & 1.10 & 1.48 & 0.35& 1.5 \\
nH100fML0.3nu1    & 100  & 0.3  & 3.58 & 3.27 & 490   & 115  & 605   & 1.10 & 1.36 & 0.09& 1.1 \\
nH100fML1nu1      & 100  & 1    & 3.39 & 3.14 & 435   & 290  & 730   & 1.09 & 1.28 & 0.07& 1.1 \\
nH100fML3nu1      & 100  & 3    & 3.03 & 2.89 & 340   & 305  & 645   & 1.09 & 1.45 & 0.52& 1.7 \\
nH100fML10nu1     & 100  & 10   & 2.33 & 2.44 & 205   & 195  & 400   & 1.00 & 1.38 & 0.90& 1.9 \\
nH100fML30nu1     & 100  & 30   & 2.55 & 2.51 & 220   & 220  & 450   & 1.03 & 1.36 & 0.57& 1.9 \\
nH100fML100nu1    & 100  & 100  & 2.55 & 2.50 & 220   & 220  & 445   & 1.02 & 1.36 & 0.70& 1.3 \\
\hline
\end{tabular}
\end{table*}

\begin{table*}
\centering
\caption[]{Physical quantities at the time of numerical shell formation ($t_{\rm SF}^{n}$; columns 4-9) and at $t=3\,t_{\rm SF}^{n}$ (columns 10-12) in models of SNRs expanding into an inhomogeneous ambient medium of intercloud density $n_{\rm H}$. The initial mass of gas in clouds is $10\times$ greater than the intercloud mass ($\nu=10$). The injection of cloud gas into the remnant occurs at a rate characterized by the value of $f_{\rm ML}$.}
\label{tab:ML_results_nu10}
\begin{tabular}{cccccccccccc}
\hline
Model & $n_{\rm H}$ & $f_{\rm ML}$ & $t_{\rm SF}^{n}$ & $r_{\rm SF}^{n}$ & $M_{\rm sw}$ & $M_{\rm inj}$ & $M_{\rm hot}$ & $p_{\rm SF}^{n}/10^{5}$ & $p_{\rm final}/10^{5}$ & $E_{\rm th}/10^{50}$ & $E_{\rm kin}/10^{50}$ \\
      & (cm$^{-3}$) &            & (kyr)           & (pc)           & ($\Msol$)   & ($\Msol$)    & ($\Msol$)   & ($\Msol\,{\rm km\,s^{-1}}$) & ($\Msol\,{\rm km\,s^{-1}}$) & (erg) & (erg)\\
\hline
nH0.01fML0.01nu10 & 0.01 & 0.01 & 690  & 166  & 6390  & 79   & 6480  & 4.20 & 5.56 & 1.8 & 1.7 \\
nH0.01fML0.03nu10 & 0.01 & 0.03 & 690  & 166  & 6380  & 240  & 6630  & 4.22 & 5.66 & 1.8 & 1.7 \\
nH0.01fML0.1nu10  & 0.01 & 0.1  & 710  & 166  & 6420  & 825  & 7260  & 4.31 & 5.86 & 1.4 & 1.4 \\
nH0.01fML0.3nu10  & 0.01 & 0.3  & 690  & 163  & 6100  & 2310 & 8420  & 4.39 & 5.68 & 1.4 & 1.1 \\
nH0.01fML1nu10    & 0.01 & 1    & 640  & 155  & 5240  & 6240 & 11500 & 4.47 & 5.56 & 2.3 & 0.66\\
nH0.01fML3nu10    & 0.01 & 3    & 550  & 140  & 3830  & 12160& 16000 & 4.24 & 5.37 & 2.7 & 0.48\\
nH0.01fML10nu10   & 0.01 & 10   & 340  & 111  & 1890  & 12320& 14200 & 3.51 & 4.07 & 1.3 & 0.54\\
nH0.01fML30nu10   & 0.01 & 30   & 230  & 85.0 & 870   & 7560 & 8420  & 3.16 & 4.36 & 0.73& 1.8 \\
nH0.01fML100nu10  & 0.01 & 100  & 135  & 61.3 & 320   & 3070 & 3400  & 2.62 & 4.21 & 2.5 & 2.5 \\
nH1fML0.01nu10    & 1    & 0.01 & 51.8 & 24.1 & 1950  & 20   & 1980  & 2.26 & 3.06 & 1.3 & 1.9 \\
nH1fML0.03nu10    & 1    & 0.03 & 51.8 & 24.0 & 1940  & 60   & 2010  & 2.26 & 3.08 & 1.1 & 1.8 \\
nH1fML0.1nu10     & 1    & 0.1  & 50.5 & 23.8 & 1890  & 190  & 2080  & 2.26 & 3.10 & 0.62& 1.7 \\
nH1fML0.3nu10     & 1    & 0.3  & 49.2 & 23.4 & 1790  & 525  & 2330  & 2.28 & 2.94 & 0.35& 1.2 \\
nH1fML1nu10       & 1    & 1    & 45.5 & 22.3 & 1550  & 1420 & 2980  & 2.27 & 2.65 & 0.54& 0.65\\
nH1fML3nu10       & 1    & 3    & 39.9 & 20.2 & 1160  & 2910 & 4050  & 2.14 & 2.31 & 0.73& 0.34\\
nH1fML10nu10      & 1    & 10   & 26.6 & 16.3 & 600   & 3420 & 3650  & 1.75 & 1.88 & 0.39& 0.44\\
nH1fML30nu10      & 1    & 30   & 18.5 & 12.7 & 290   & 2430 & 2690  & 1.59 & 1.86 & 0.18& 1.1 \\
nH1fML100nu10     & 1    & 100  & 11.5 & 9.2  & 110   & 1020 & 1140  & 1.42 & 2.26 & 1.8 & 2.2 \\
nH100fML0.01nu10  & 100  & 0.01 & 3.67 & 3.33 & 515   & 4.1  & 525   & 1.10 & 1.48 & 0.76& 1.8 \\
nH100fML0.03nu10  & 100  & 0.03 & 3.67 & 3.33 & 515   & 12.4 & 535   & 1.10 & 1.49 & 0.73& 1.7 \\
nH100fML0.1nu10   & 100  & 0.1  & 3.58 & 3.29 & 500   & 39   & 545   & 1.10 & 1.48 & 0.35& 1.5 \\
nH100fML0.3nu10   & 100  & 0.3  & 3.58 & 3.27 & 490   & 115  & 605   & 1.10 & 1.36 & 0.09& 1.1 \\
nH100fML1nu10     & 100  & 1    & 3.39 & 3.14 & 435   & 325  & 755   & 1.09 & 1.20 & 0.14& 0.69\\
nH100fML3nu10     & 100  & 3    & 3.03 & 2.89 & 340   & 700  & 945   & 1.06 & 1.10 & 0.22& 0.36\\
nH100fML10nu10    & 100  & 10   & 2.12 & 2.36 & 185   & 890  & 845   & 0.95 & 0.97 & 0.12& 0.40\\
nH100fML30nu10    & 100  & 30   & 1.51 & 1.86 & 90    & 720  & 740   & 0.88 & 0.90 & 0.05& 0.85\\
nH100fML100nu10   & 100  & 100  & 0.96 & 1.35 & 34    & 315  & 360   & 0.81 & 1.13 & 0.70& 1.9 \\
\hline
\end{tabular}
\end{table*}

\begin{table*}
\centering
\caption[]{Physical quantities at the time of numerical shell formation ($t_{\rm SF}^{n}$; columns 4-9) and at $t=3\,t_{\rm SF}^{n}$ (columns 10-12) in models of SNRs expanding into an inhomogeneous ambient medium of intercloud density $n_{\rm H}$. There is effectively an infinite reservoir of cloud mass ($\nu=10^{8}$). The injection of cloud gas into the remnant occurs at a rate characterized by the value of $f_{\rm ML}$.}
\label{tab:ML_results}
\begin{tabular}{cccccccccccc}
\hline
Model & $n_{\rm H}$ & $f_{\rm ML}$ & $t_{\rm SF}^{n}$ & $r_{\rm SF}^{n}$ & $M_{\rm sw}$ & $M_{\rm inj}$ & $M_{\rm hot}$ & $p_{\rm SF}^{n}/10^{5}$ & $p_{\rm final}/10^{5}$ & $E_{\rm th}/10^{50}$ & $E_{\rm kin}/10^{50}$ \\
      & (cm$^{-3}$) &            & (kyr)           & (pc)           & ($\Msol$)   & ($\Msol$)    & ($\Msol$)   & ($\Msol\,{\rm km\,s^{-1}}$) & ($\Msol\,{\rm km\,s^{-1}}$) & (erg) & (erg) \\
\hline
nH0.01fML0.01nu1e8& 0.01 & 0.01& 690  & 166  & 6400 & 79    & 6480  & 4.20 & 5.56 & 1.8 & 1.7 \\
nH0.01fML0.03nu1e8& 0.01 & 0.03& 690  & 166  & 6400 & 240   & 6630  & 4.22 & 5.66 & 1.8 & 1.7 \\
nH0.01fML0.1nu1e8 & 0.01 & 0.1 & 710  & 166  & 6420 & 825   & 7260  & 4.31 & 5.84 & 1.5 & 1.5 \\
nH0.01fML0.3nu1e8 & 0.01 & 0.3 & 690  & 163  & 6100 & 2300  & 8420  & 4.39 & 5.68 & 1.4 & 1.1 \\	
nH0.01fML1nu1e8   & 0.01 & 1   & 640  & 155  & 5240 & 6240  & 11500 & 4.47 & 5.56 & 2.3 & 0.66\\
nH0.01fML3nu1e8   & 0.01 & 3   & 550  & 140  & 3830 & 12170 & 16000 & 4.24 & 5.55 & 3.5 & 0.43\\
nH0.01fML10nu1e8  & 0.01 & 10  & 340  & 110  & 1890 & 12770 & 14700 & 3.50 & 4.59 & 3.2 & 0.31\\
nH0.01fML30nu1e8  & 0.01 & 30  & 230  & 85.2 & 870  & 12340 & 12820 & 2.90 & 3.79 & 2.8 & 0.24\\ 
nH0.01fML100nu1e8 & 0.01 & 100 & 137  & 61.3 & 320  & 9770  & 9840  & 2.45 & 3.03 & 2.0 & 0.23\\
nH1fML0.01nu1e8   & 1    & 0.01& 51.8 & 24.1 & 1950 & 20    & 1980  & 2.26 & 3.06 & 1.3 & 1.9 \\
nH1fML0.03nu1e8   & 1    & 0.03& 51.8 & 24.0 & 1940 & 60    & 2010  & 2.26 & 3.08 & 1.1 & 1.8 \\
nH1fML0.1nu1e8    & 1    & 0.1 & 50.5 & 23.8 & 1885 & 190   & 2080  & 2.26 & 3.10 & 0.70& 1.7 \\
nH1fML0.3nu1e8    & 1    & 0.3 & 49.2 & 23.4 & 1790 & 525   & 2330  & 2.28 & 2.94 & 0.35& 1.2 \\
nH1fML1nu1e8      & 1    & 1   & 45.5 & 22.3 & 1550 & 1420  & 2980  & 2.27 & 2.66 & 0.55& 0.63\\
nH1fML3nu1e8      & 1    & 3   & 39.9 & 20.2 & 1160 & 2910  & 4050  & 2.14 & 2.32 & 0.87& 0.32\\
nH1fML10nu1e8     & 1    & 10  & 26.6 & 16.3 & 605  & 3440  & 3650  & 1.75 & 1.94 & 0.89& 0.23\\
nH1fML30nu1e8     & 1    & 30  & 17.2 & 12.5 & 275  & 3180  & 3090  & 1.50 & 1.64 & 0.74& 0.20\\
nH1fML100nu1e8    & 1    & 100 & 11.5 & 9.20 & 110  & 2960  & 2550  & 1.33 & 1.43 & 0.63& 0.18\\
nH100fML0.01nu1e8 & 100  & 0.01& 3.68 & 3.33 & 515  & 4.1   & 525   & 1.10 & 1.48 & 0.76& 1.8 \\
nH100fML0.03nu1e8 & 100  & 0.03& 3.68 & 3.33 & 515  & 12    & 535   & 1.10 & 1.49 & 0.73& 1.7 \\
nH100fML0.1nu1e8  & 100  & 0.1 & 3.58 & 3.29 & 500  & 39    & 545   & 1.10 & 1.48 & 0.35& 1.5 \\
nH100fML0.3nu1e8  & 100  & 0.3 & 3.58 & 3.27 & 490  & 115   & 605   & 1.10 & 1.36 & 0.09& 1.1 \\
nH100fML1nu1e8    & 100  & 1   & 3.39 & 3.14 & 435  & 330   & 750   & 1.09 & 1.20 & 0.14& 0.69\\
nH100fML3nu1e8    & 100  & 3   & 3.03 & 2.89 & 340  & 700   & 945   & 1.06 & 1.10 & 0.23& 0.36\\
nH100fML10nu1e8   & 100  & 10  & 2.17 & 2.36 & 185  & 890   & 840   & 0.95 & 0.98 & 0.26& 0.22\\
nH100fML30nu1e8   & 100  & 30  & 1.45 & 1.84 & 87   & 900   & 730   & 0.86 & 0.88 & 0.23& 0.19\\
nH100fML100nu1e8  & 100  & 100 & 0.96 & 1.35 & 34   & 810   & 595   & 0.78 & 0.80 & 0.19& 0.18\\  
\hline
\end{tabular}
\end{table*}

\begin{table*}
\centering
\caption[]{Physical quantities at $t=2$\,Myr in models of SNRs created by clustered SN explosions. The remnant expands into either a homogeneous ($f_{\rm ML}=0$) or inhomogeneous ambient medium, experiencing mass-loading from engulfed clouds in the latter case. The initial mass of gas in clouds is $\nu$ times greater than the intercloud mass. The injection of cloud gas occurs at a rate characterized by the value of $f_{\rm ML}$. See Sec.~\ref{sec:multSNenoML} for further details.}
\label{tab:mSNe_results}
\begin{tabular}{cccccccccccc}
\hline
Model & \#SNe & $f_{\rm ML}$ & $\nu$ & $r_{\rm snr}$ & $M_{\rm sw}$ & $M_{\rm inj}$ & $M_{\rm hot}$ & $p/10^{6}$ & $E_{\rm th}$ & $E_{\rm kin}$ & $E_{\rm tot}$\\
      &      &             &       & (pc)         & ($10^{5}\,\Msol$)  & ($10^{5}\,\Msol$)    & ($\Msol$)   & ($\Msol\,{\rm km\,s^{-1}}$) & ($10^{51}$\,erg) & ($10^{51}$\,erg) & ($10^{51}$\,erg)\\
\hline
nH1fML0mSNe      & 10 & 0 & -        & 123.4 & 2.63 & 0   & 207    & 6.1  & 2.81 & 1.42 & 4.23 \\
nH1fML1nu1e8mSNe & 10 & 1 & $10^{8}$ & 70.7  & 0.50 & 16.1 & 0.0    & 9.71 & 6.83 & 0.75 & 7.59 \\
nH1fML10nu1e8mSNe& 10 & 10& $10^{8}$ & 63.3  & 0.36 & 99.5 & 0.0    & 53.2 & 41.1 & 3.34 & 44.5 \\
nH1fML1nu10mSNe  & 10 & 1 & 10       & 80.9 & 0.74 & 6.64 & 94.8    & 8.66 & 3.83 & 1.04 & 4.87 \\
nH1fML0E52       & 1  & 0 & -        & 124.1 & 2.68 & 0   & 605    & 3.0  & 1.30 & 0.39 & 1.69 \\
\hline
\end{tabular}
\end{table*}

\begin{figure*}
\includegraphics[width=17.5cm]{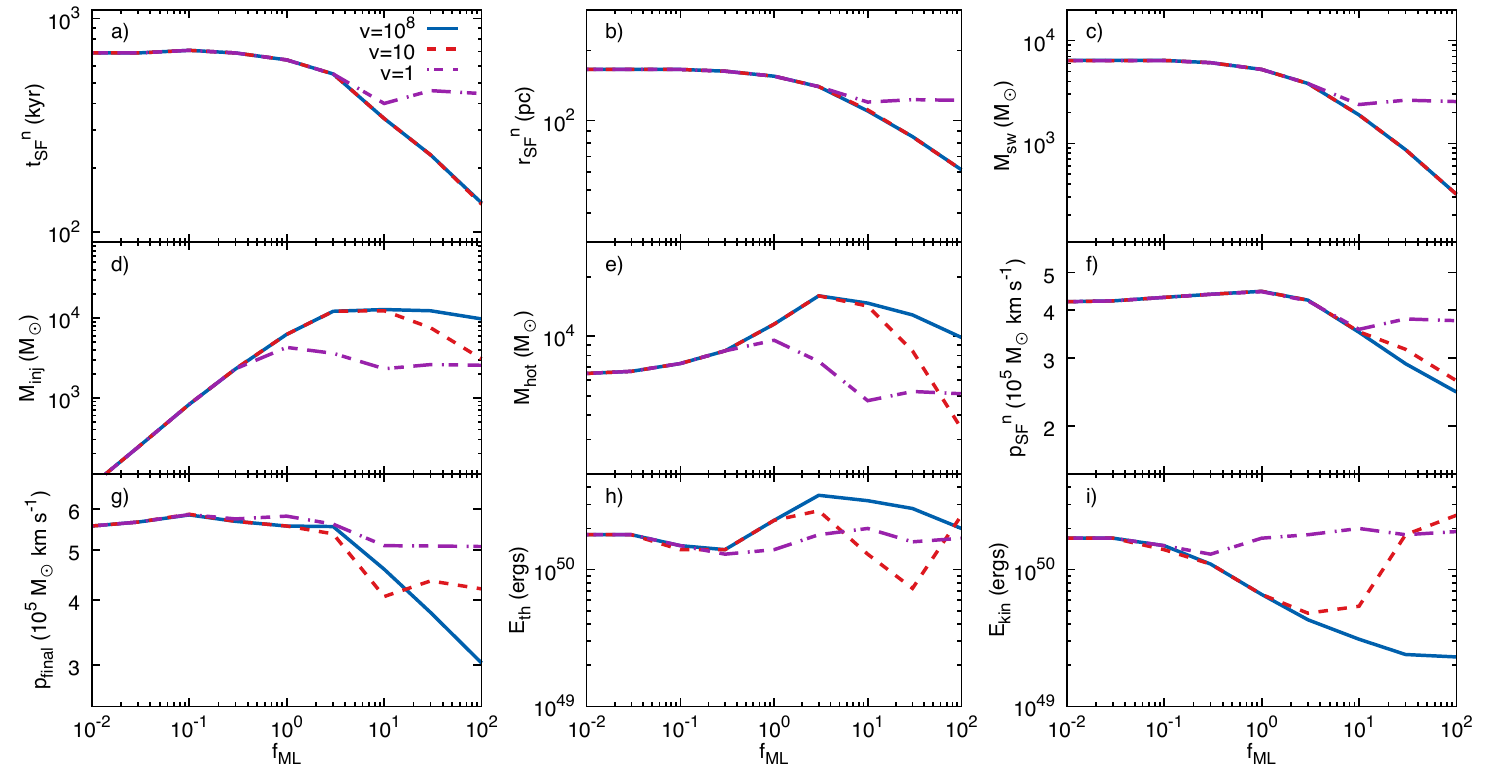}
\caption{Various statistics as a function of $f_{\rm ML}$ and $\nu$ for simulations with $n_{\rm H} =0.01\,{\rm cm^{-3}}$. Panels a)-f) are for quantities evaluated at the time of shell formation, $t_{\rm SF}$: a) the age of the remnant; b) its radius; c) the mass of swept-up intercloud gas; d) the mass of injected gas; e) the mass of hot gas; f) the radial momentum. Panels g)-i) are for quantities evaluated at $t = 3\,t_{\rm SF}$: g) the ``final'' radial momentum; h) the thermal energy; i) the kinetic energy.}
\label{fig:nH0.01varnu_stats}
\end{figure*}

\begin{figure*}
\includegraphics[width=17.5cm]{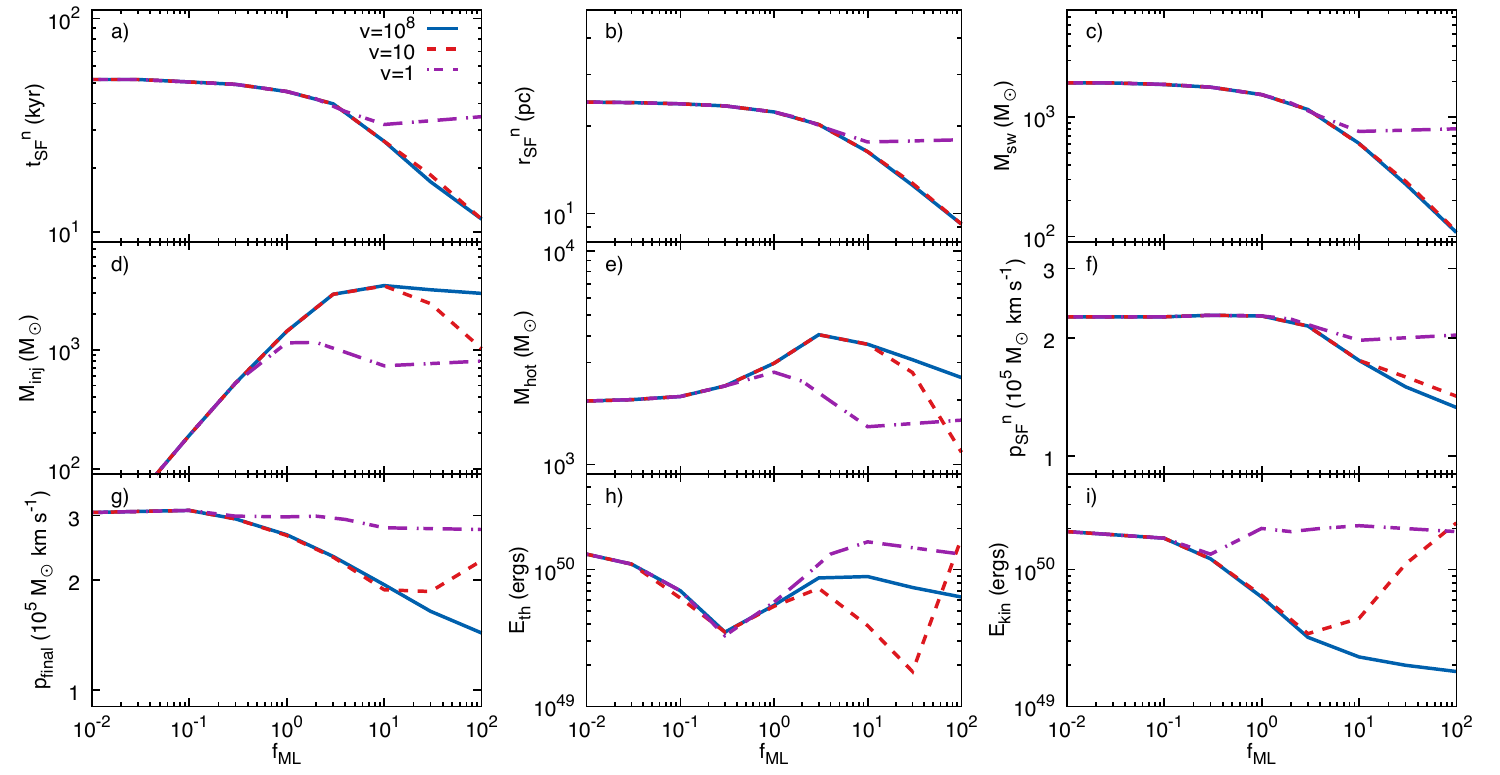}
\caption{Various statistics as a function of $f_{\rm ML}$ and $\nu$ for simulations with $n_{\rm H} =1\,{\rm cm^{-3}}$. Panels a)-f) are for quantities evaluated at the time of shell formation, $t_{\rm SF}$: a) the age of the remnant; b) its radius; c) the mass of swept-up intercloud gas; d) the mass of injected gas; e) the mass of hot gas; f) the radial momentum. Panels g)-i) are for quantities evaluated at $t = 3\,t_{\rm SF}$: g) the ``final'' radial momentum; h) the thermal energy; i) the kinetic energy.}
\label{fig:nH1varnu_stats}
\end{figure*}

\begin{figure*}
\includegraphics[width=17.5cm]{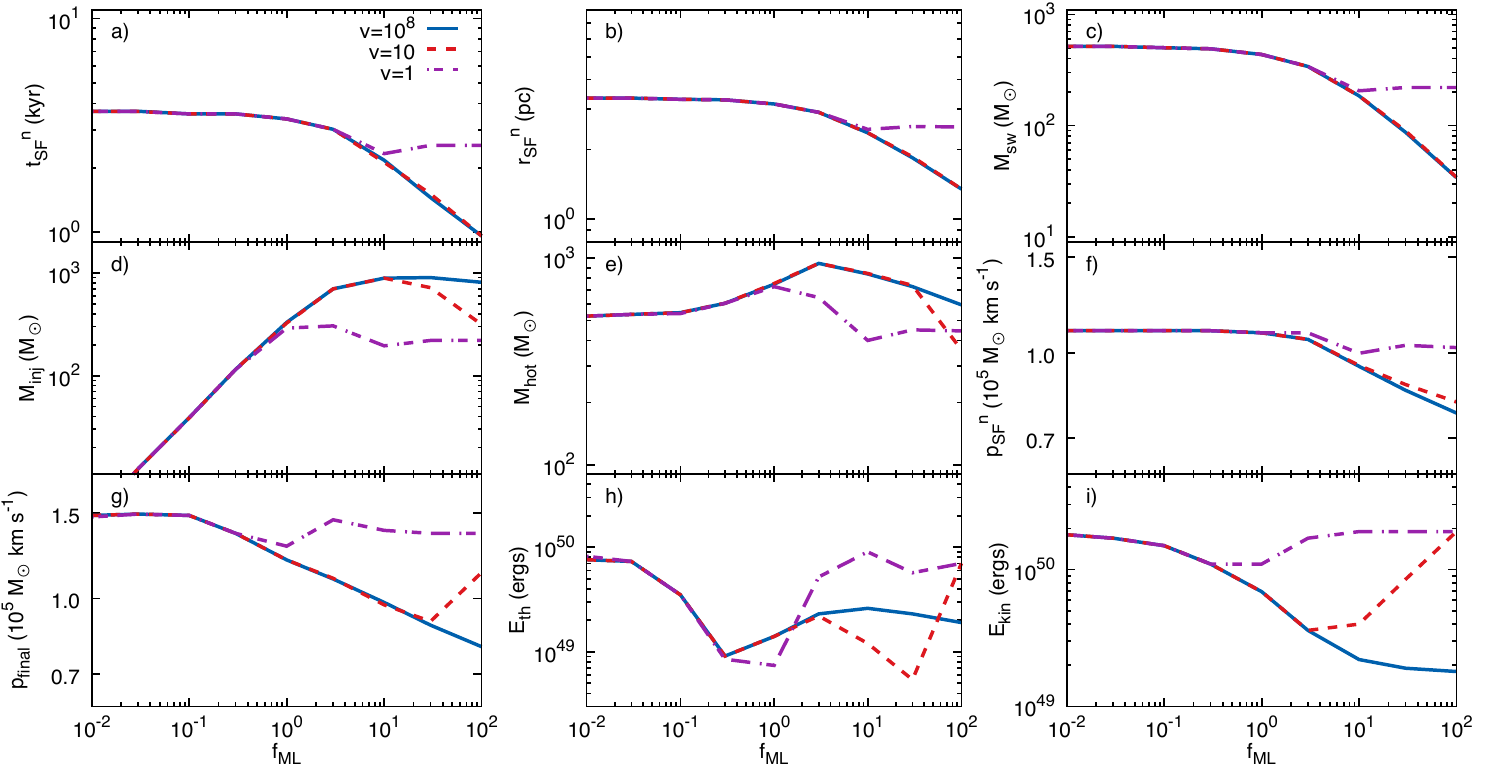}
\caption{Various statistics as a function of $f_{\rm ML}$ and $\nu$ for simulations with $n_{\rm H} =100\,{\rm cm^{-3}}$. Panels a)-f) are for quantities evaluated at the time of shell formation, $t_{\rm SF}$: a) the age of the remnant; b) its radius; c) the mass of swept-up intercloud gas; d) the mass of injected gas; e) the mass of hot gas; f) the radial momentum. Panels g)-i) are for quantities evaluated at $t = 3\,t_{\rm SF}$: g) the ``final'' radial momentum; h) the thermal energy; i) the kinetic energy.}
\label{fig:nH100varnu_stats}
\end{figure*}

\section{Dependence on the floor temperature}
\label{sec:tfloor}
\begin{figure*}
\includegraphics[width=17.5cm]{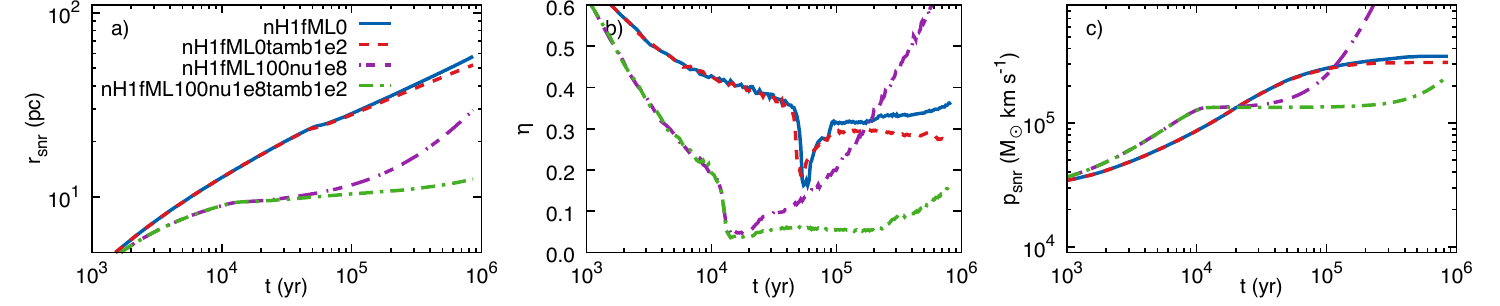}
\caption{The time evolution of models nH1fML0 (no mass-loading) and nH1fML100nu1e8 with $T_{\rm floor} = T_{\rm amb} = 10^{4}$\,K and with $T_{\rm floor} = T_{\rm amb} = 10^{2}$\,K. The panels show: a) the radius; b) the deceleration parameter; c) the total radial momentum.}
\label{fig:nH1variousfMLnu1e8vartamb_stats}
\end{figure*}

In this appendix we examine how the floor temperature, $T_{\rm floor}$, and the temperature of the ambient medium, $T_{\rm amb}$, affects our simulations. In general these temperatures may differ but in this work we set $T_{\rm floor} = T_{\rm amb} = 10^{4}$\,K. As we have noted in the main text, the late-time evolution of the remnant can be affected by the value of $T_{\rm floor}$ or $T_{\rm amb}$. Fig.~\ref{fig:nH1variousfMLnu1e8vartamb_stats} shows the radius, deceleration parameter, and the total radial momentum as a function of time for models with and without mass-loading, with $T_{\rm floor} = T_{\rm amb} = 10^{4}$\,K and with $T_{\rm floor} = T_{\rm amb} = 10^{2}$\,K. The first thing to note is that the value of $T_{\rm floor/amb}$ also affects simulations without mass-loading. A lower value for $T_{\rm floor/amb}$ results in a slightly reduced remnant radius at late times, and a deceleration parameter that continues to decline towards the PDS value. The total radial momentum at late times is also reduced slightly, by 12 per cent. Hence the value of $p_{\rm final}$ (evaluated at $t=10\,t_{\rm SF}$ in simulations without mass-loading) is weakly dependent on the value of $T_{\rm floor/amb}$.

The value of $T_{\rm floor/amb}$ has a much greater impact in the simulations with strong mass-loading, where the late-time behaviour of simulations with $T_{\rm floor/amb}=10^{4}$\,K is dominated by the very significant amount of thermal energy that is effectively added with the injected mass. This prevents the remnant from slowing below an expansion speed of $\approx 20\kmps$ (see Sec.~\ref{sec:variousfMLinfiniteReservoir}) and causes steeply rising trajectories for $r_{\rm snr}$, $\eta$, and $p_{\rm snr}$. Setting $T_{\rm floor/amb}=10^{2}$\,K reduces the additional thermal energy added into the remnant by a factor of 100, delaying and reducing its late-time effect. By evaluating $p_{\rm final}$ at $t=3\,t_{\rm SF}$ in simulations with mass-loading we find that its value is largely insensitive to $T_{\rm floor/amb}$.

\bsp	
\label{lastpage}
\end{document}